\documentclass[aps,prd,10pt,superscriptaddress,twocolumn,nofootinbib,dvipsnames,longbibliography]{revtex4-1}

\usepackage[pdftex]{color}
\usepackage[sort&compress]{natbib}
\usepackage[colorlinks=true,linkcolor=purple,filecolor=orange,urlcolor=blue,citecolor=purple,pdftex,plainpages=false]{hyperref}
\usepackage{url}
\pdfoutput=1
\usepackage{ifpdf}
\usepackage[utf8]{inputenc}
\usepackage{color,hhline,psfrag,rotating}
\usepackage{dcolumn}
\usepackage{verbatim}
\usepackage{lipsum}
\usepackage{bm}
\usepackage{slashed}
\usepackage{braket}
\usepackage{rotating}

\usepackage{graphicx}
\usepackage[dvipsnames]{xcolor}
\usepackage{amsfonts,amssymb,amsmath}
\usepackage{booktabs}  

\newcommand\brachat[1]{\mathord{\mathop{#1}\limits^{\!\!\scriptscriptstyle(\wedge)}}}

\mathchardef\mhyphen="2D

\begin{document}

\global\let\newpage\relax

\title{Toward accurate form factors for $B$-to-light meson decay from lattice QCD}

\author{W.~G.~Parrott}
\email[]{w.parrott.1@research.gla.ac.uk}
\affiliation{SUPA, School of Physics and Astronomy, University of Glasgow, Glasgow, G12 8QQ, United Kingdom}
\author{C.~Bouchard}
\email[]{chris.bouchard@glasgow.ac.uk}
\affiliation{SUPA, School of Physics and Astronomy, University of Glasgow, Glasgow, G12 8QQ, United Kingdom}
\author{C.~T.~H.~Davies}
\email[]{christine.davies@glasgow.ac.uk}
\affiliation{SUPA, School of Physics and Astronomy, University of Glasgow, Glasgow, G12 8QQ, United Kingdom}
\author{D.~Hatton}
\email[]{d.hatton.1@research.gla.ac.uk}
\affiliation{SUPA, School of Physics and Astronomy, University of Glasgow, Glasgow, G12 8QQ, United Kingdom}
\collaboration{HPQCD Collaboration}
\homepage{http://www.physics.gla.ac.uk/HPQCD}
\noaffiliation

\date{\today}

\begin{abstract}
  We present the results of a lattice QCD calculation of the scalar and vector form factors for the unphysical $B_s\to\eta_s$ decay, over the full physical range of $q^2$. 
  This is a useful testing ground both for lattice QCD and for our wider understanding of the behaviour of form factors. 
Calculations were performed using the highly improved staggered quark (HISQ) action on $N_f = 2 + 1 + 1$ gluon ensembles generated by the MILC Collaboration with an improved gluon action and HISQ sea quarks. 
We use three lattice spacings and a range of heavy quark masses from that of charm to bottom, all in the HISQ formalism. 
This permits an extrapolation in the heavy quark mass and lattice spacing to the physical point and nonperturbative renormalisation of the vector matrix element on the lattice. 
We find results in good agreement with previous work using nonrelativistic QCD $b$ quarks and with reduced errors at low $q^2$, supporting the effectiveness of our heavy HISQ technique as a method for calculating form factors involving heavy quarks. 
A comparison with results for other decays related by SU(3) flavour symmetry shows that the impact of changing the light daughter quark is substantial but changing the spectator quark has very little effect. 
We also map out form factor shape parameters as a function of heavy quark mass and compare to heavy quark effective theory expectations for mass scaling at low and high recoil. 
This work represents an important step in the progression from previous work on heavy-to-heavy decays ($b\to c$) to the numerically more 
challenging heavy-to-light decays.
\end{abstract}

\maketitle

\section{Introduction} \label{sec:intro}
Determinations of form factors for weak semileptonic meson decays can be combined with experimental results to provide important tests of the Standard Model (SM). 
Decays of $b$ quarks are of particular interest as they allow determination of some of the least well-known elements of the Cabibbo–Kobayashi–Maskawa (CKM) matrix~\cite{PhysRevLett.10.531,10.1143/PTP.49.652} and tests 
of the unitarity of that matrix, a foundation of the weak sector of the SM. 
Increasingly small experimental uncertainties in CKM-dependent decay rates must be met with precise determinations of form factors from the theoretical side to pin down the CKM matrix elements (see, for example,~\cite{2019arXiv190901213L,2019arXiv190809713B}). 
The shape of the differential decay rate in $q^2$, the squared momentum transfer between the initial and final states, parameterised by the form factors, provides added detail when testing the SM. 
Lattice quantum chromodynamics (lattice QCD) is the only model-independent
method for calculating the hadronic form factors for such 
decays and has been used successfully for many such calculations. 
For a review, see~\cite{Aoki:2019cca}.

Resolving the $b$ quark on the lattice requires a sufficiently small lattice spacing, $a < 1/m_b \sim 0.05$\,fm.
This means that lattice QCD calculations can currently only reach the $b$ quark mass on the finest lattices available.

One approach to address this difficulty relies on the use of an effective theory description of the $b$ quark.
Examples include the relativistic heavy quark action~\cite{Flynn:2015mha}, the Fermilab action~\cite{Lattice:2015tia, Bazavov:2019aom}, heavy quark effective theory (HQET)~\cite{Bahr:2019eom}, and nonrelativistic QCD (NRQCD)~\cite{Bouchard:2013eph, Horgan:2013pva, Horgan:2013hoa}.
Each of these must match the relevant effective theory to QCD and therefore suffer from associated matching errors.  
For the case of NRQCD, such matching errors are a dominant source of uncertainty.

Alternatively, with an action sufficiently improved to reduce heavy quark discretisation effects, one can avoid this use of effective theory and simulate over a range of heavy quark masses $m_h \lesssim m_b$ and then extrapolate (or interpolate if, for example, static quark results are available) to $m_b$.
Examples of this approach include the ratio method using the twisted mass formulation~\cite{Blossier:2009hg, Bussone:2016iua}, application of the M\"obius domain wall formulation to the $b$ quark~\cite{Colquhoun:2017gfi} and our recent works using the highly improved staggered quark (HISQ) action for the $b$ quark in several $b\to c$ decays~\cite{McLean:2019qcx,Harrison:2020gvo, Harrison:2020nrv}.

The HISQ action~\cite{2007PhRvD..75e4502F} provides an accurate discretisation of the Dirac equation for relatively heavy quarks~\cite{McNeile:2011ng}. 
It allows us to normalise lattice currents nonperturbatively using conserved currents, avoiding sizable systematic errors from perturbative truncation in the renormalisation factors for the nonrelativistic case. 
This ``heavy-HISQ" approach must be carried out on fine lattices, with $a < 0.1$~fm so that $am_h$ is not too large. 
On our finest lattices, $am_b < 1$. 
In practice, we work at several values of $a$ and of the heavy quark mass so that we can map out both discretisation effects and physical dependence on the heavy quark mass to determine the result at $m_b$ and in the continuum. 
A further advantage of working on such fine lattices is that we can reach higher physical values of momentum transfer as the lattice spacing gets smaller. 
This is particularly important for $b$ decays where the $q^2$ range for the decay is large. 
With the heavy-HISQ approach the range of accessible $q^2$ values grows on finer lattices in step with the range of heavy quark masses. 
This means that we can cover the full $q^2$ range of the heavy quark decay all the way up to that of the $b$~\cite{McLean:2019qcx}. 

The end game of this program is the determination of form factors for transitions that involve physical $u$ and $d$ quarks, such as $B \rightarrow \pi$.
In this work, we take an important step in extending our use of the HISQ action for the $b$ quark in $b\to c$ decays toward the more demanding $b\to u,d$ decays by studying the $b\to s$ transition.
As $m_s \ll m_c$, this allows us to gauge the success of this approach for $b$-to-light form factors while benefiting from both a significant savings in computational cost and the typically less noisy correlators associated with $s$ quarks.
Fixing the daughter quark to the strange quark mass on each ensemble removes the need to perform a chiral extrapolation, thereby simplifying the continuum extrapolation, a key component in our study of the efficacy of the heavy-HISQ approach.
Here we study the $B_s\to\eta_s$ decay, where the $\eta_s$ is an unphysical $s\overline{s}$ pseudoscalar meson--an easier to analyse, cheaper to compute substitute for a pion, with the same quantum numbers and no valence annihilation.
For the purposes of assessing the viability of this approach, we focus on the scalar and vector form factors.\footnote{The flavor-changing neutral current responsible for the $b\to s$ decay in the SM would also involve the tensor form factor.  The tensor form factor is typically noisier, so we ignore it here.  We also ignore any difficulties associated with converting form factors into decay rates, such as the $c\bar c$ resonances that appear in the phenomenology of $B\to K\mu\bar\mu$.}

The form factors should not be greatly affected by changing the spectator quark from an $s$ quark to a $u/d$ quark, so studying this decay provides an estimate of the level of precision achievable in the computationally more expensive $B\to K$ form factor calculation. 
The heavy-HISQ approach allows us to extract the dependence of the form factors on the heavy quark mass as it varies from $m_c$ to $m_b$, permitting useful tests for expectations from heavy quark symmetry. 

The paper is laid out as follows. 
In Sec.~\ref{sec:calc} we set out the details of our lattice QCD calculation, including analysis of the correlation functions, normalisation of the lattice currents and our fits to the form factors enabling results to be obtained for $B_s \rightarrow \eta_s$ decay in the continuum limit. 
Sec.~\ref{sec:results} gives results and compares them both to expectations from heavy quark symmetry and to previous lattice QCD results for decay processes connected to $B_s \rightarrow \eta_s$ and $D_s \rightarrow \eta_s$ by SU(3) flavour symmetry, either for the active light quark in the decay or the spectator light quark. 
Finally, Sec.~\ref{sec:conclusions} gives our conclusions. 

\section{Calculation Details}\label{sec:calc}
\subsection{Form factors}\label{Sec:formfactors}
The aim of our calculation is to determine the matrix element for the $V-A$ electroweak current between $B_s$ and $\eta_s$ mesons, $\bra{B_s}V^{\mu}-A^{\mu}\ket{\eta_s}$. 
Here the vector current is defined as $V^{\mu}=\bar{\psi}_b\gamma^{\mu}\psi_s$ and the axial vector current is $A^{\mu}=\bar{\psi}_b\gamma^5\gamma^{\mu}\psi_s$. 
For pseudoscalar to pseudoscalar decays, only contributions from the vector part of the $V-A$ current are present, as a result of QCD parity invariance. 

Our heavy-HISQ approach works by determining the $B_s$ meson matrix elements from a set of matrix elements for mesons in which the $b$ quark is replaced by a heavy quark with mass $m_h < m_b$. 
We denote these pseudoscalar heavy-strange mesons generically by $H_s$. 
The form factors $f_+(q^2)$ and $f_0(q^2)$ that are determined from the matrix elements are a function of $q^2 = (p_{H_s} - p_{\eta_s})^2$, and we compute these across the full kinematic range, $0\leq q^2\leq q_{\rm max}^2 = (M_{H_s} - M_{\eta_s})^2$. 
As $m_h \rightarrow m_b$ this becomes the full range for the $B_s$ decay. 

The connection between the matrix elements of the lattice temporal vector and scalar currents and the form factors of interest, $f_+(q^2)$ and $f_0(q^2)$, is
\begin{align}
  &Z^0_V Z_{\rm disc} \bra{\eta_s} V^{0} \ket{\widehat{H}_s} = \nonumber \\
  &f_+^{H_s\to\eta_s}(q^2) \Big( E_{H_s} + E_{\eta_s} - \frac{M_{H_s}^2-M_{\eta_s}^2}{q^2} (E_{H_s} - E_{\eta_s}) \Big)\nonumber \\
    &+f_0^{H_s\to\eta_s}(q^2)\frac{M_{H_s}^2-M_{\eta_s}^2}{q^2} (E_{H_s} - E_{\eta_s}),
\label{Eq:vec}
\end{align}
\begin{align}
 &Z_{\rm disc}\bra{\eta_s}S\ket{H_s}=\frac{M_{H_s}^2-M_{\eta_s}^2}{m_h-m_s}f_0^{H_s\to\eta_s}(q^2).
  \label{Eq:sca}
\end{align}
Bilinears constructed from staggered quarks have a ``taste" degree of freedom and, as will be discussed below, we need to arrange the tastes of mesons and lattice currents appropriately so that tastes cancel in the calculated correlation functions. 
Here, in spin-taste notation~\cite{2007PhRvD..75e4502F}, the lattice currents are $S = \bar{\psi}_s 1 \otimes1 \psi_b$ and $V^{0} = \bar{\psi}_s \gamma^{0} \otimes \xi^0 \psi_b$ and $H_s$ and $\widehat{H}_s$ denote Goldstone and local non-Goldstone heavy-strange pseudoscalar mesons, respectively. 
Eq.~(\ref{Eq:sca}) comes from the partially conserved vector current (PCVC) relation~\cite{Na:2010uf}, which also leads to the renormalisation of the vector matrix element~\cite{Na:2010uf, Koponen:2013tua} (see Sec.~\ref{sec:norm}).

We also require that the matrix element is analytic as $q^2\to 0$. 
We can see from Eq.~(\ref{Eq:vec}) that this demands 
\begin{equation}
  f_+^{H_s \to \eta_s}(0) = f_0^{H_s \to \eta_s}(0), 
\end{equation}
where we will drop the superscript from now on.

Both matrix elements are calculated using a Goldstone pseudoscalar strange-strange $\eta_s$ bilinear, $\eta_s = \bar{\psi}_s \gamma^5 \otimes \xi^5 \psi_s$, whilst the scalar uses the Goldstone pseudoscalar heavy-strange $H_s = \bar{\psi}_b \gamma^5 \otimes \xi^5 \psi_s$, and the vector uses the non-Goldstone pseudoscalar heavy-strange $\widehat{H}_s = \bar{\psi}_b \gamma^5 \gamma^0 \otimes \xi^5 \xi^0 \psi_s$. 
All of these operators are local, giving less noisy correlation functions than their point-split counterparts. 

\subsection{Lattice details}\label{sec:latt}
The calculation was run on ensembles of gluon field configurations generated by MILC~\cite{PhysRevD.82.074501,PhysRevD.87.054505}.
These include in the sea two degenerate light quarks, strange and charm quarks, with masses $m_l^{\rm sea}$, $m_s^{\rm sea}$, and $m_c^{\rm sea}$, respectively, using the HISQ action. 
The three ensembles used have parameters listed in Table~\ref{tab:ensembles}. 
The gluon action is Symanzik improved to remove discretisation errors through $\mathcal{O}(\alpha_sa^2)$~\cite{PhysRevD.59.074502}. 
Our calculation follows the approach in the calculation of $B_s\to D_s$ in~\cite{McLean:2019qcx} but with a strange daughter quark in lieu of a charm. 
The ensembles that we use here have unphysically heavy light quark masses (of value around 1/5 of the $s$ quark mass). 
In~\cite{McLean:2019qcx}, little effect was seen on the form factors from the light quark mass in the sea. 
We similarly expect little effect here since $B_s \rightarrow\eta_s$ does not involve any valence light quarks. 
Our main focus here is to test the heavy quark mass dependence and so we simply address the mistuning of sea light quark masses when we extrapolate to the physical point in Sec.~\ref{sec:extrap}.

\begin{table}[t]
  \caption{Gluon field ensembles used in this work. The Wilson flow parameter, $w_0=0.1715(9)\,\text{fm}$, is determined in~\cite{Dowdall:2013rya}, following the approach outlined in~\cite{Borsanyi:2012zs}, and is used to calculate the lattice spacing $a$ via values for $w_0/a$, in column 3, which are from~\cite{McLean:2019qcx}. Column 4 gives the spatial ($N_s$) and temporal ($N_t$) dimensions of each lattice in lattice units, whilst columns 5--7 give the masses of the sea quarks. }
  \begin{center}
    \begin{tabular}{c c c c c c c c c c}
      \hline
      Set & Handle & $w_0/a$ & $N_s^3\times N_t$ & $am^{\text{sea}}_{l}$ & $am^{\text{sea}}_{s}$ & $am^{\text{sea}}_{c}$  \\ [0.5ex]
      \hline
      1 & Fine & 1.9006(20) & $32^3\times96$ & 0.0074 & 0.037 & 0.440 \\ [1ex]
      2 & Superfine & 2.896(6) & $48^3\times144$ & 0.0048 & 0.024 & 0.286 \\ [1ex]
      3 & Ultrafine & 3.892(12) &  $64^3\times192$ & 0.00316 & 0.0158 & 0.188  \\ [1ex]
      \hline
    \end{tabular}
  \end{center}
  \label{tab:ensembles}
\end{table}

We denote the heavy quark $h$ and its mass $m^{\rm val}_h$ and use a range of heavy masses from the physical charm to $am_h^{\rm val}=0.8$, the point where discretisation errors start to become significant, on each set of gluon configurations. 
This allows us to perform a fit to our results as a function of heavy quark mass and obtain results at the physical $b$ mass. 
At the same time we determine the dependence of the form factors on the heavy mass from the charm to the bottom with  $D_s \to \eta_s$ and $B_s \rightarrow \eta_s$ at the two ends of the range. 
On the finest lattice $am_h^{\rm val} = 0.8$ is close to the physical $b$ mass, allowing good control of the subsequent extrapolation to $m_b$.

We choose a range of daughter momenta so as to give good coverage of the full momentum transfer range of the decay (see Table~\ref{tab:simulation}) and implement these momenta using twisted boundary conditions on the daughter strange quark in the $\eta_s$, as described in~\cite{Guadagnoli:2005be}. 
The heavy meson remains at rest in all stages of the calculation, meaning the strange spectator and heavy quark have no twist applied.

\begin{table}[t]
  \caption{
Values of simulation parameters on each ensemble used in this work.
Valence strange quark masses $am_s^{\rm val}$ are tuned as described in~\cite{Chakraborty:2014aca}.
Valence heavy quark masses $am_h^{\rm val}$ are chosen to span the range from the physical charm, tuned as in~\cite{Chakraborty:2014aca}, to $am_h^{\rm val} = 0.8$.
Simulated $\eta_s$ momenta $a\vec{p}_{\eta_s}$ are fixed using twisted boundary conditions as described in the text.
On each ensemble, we use $n_{\rm cfg}$ configurations and $n_{\rm src}$ time sources.
Data are generated for multiple temporal source-sink separations $T$ between the $\eta_s$ and $H_s$ mesons.
}
  \begin{center} 
    \begin{tabular}{c c c c}
      \hline
       & Set 1  & Set 2 & Set 3\\ [0.5ex]
      \hline
      $am_{s}^{\text{val}}$ & 0.0376 & 0.0234&0.0165\\ [1ex]
      \hline
      
      $am^{\text{val}}_{h}$ &0.449&0.274&0.194\\ [1ex]
                          & 0.566 &0.45& 0.45 \\ [1ex]
                          &0.683&0.6&0.6\\ [1ex]
                          & 0.8 & 0.8&0.8 \\ [1ex]

      \hline
      $|a{\vec{p}}_{\eta_s}|$ &0&0&0 \\ [1ex]
      &0.0728 &0.1430 &0.0600 \\ [1ex]
      &0.2180 &0.2390 & 0.1300\\ [1ex]
       &0.3641 & 0.3340&0.1900  \\ [1ex]
      &0.4370 & 0.4108&0.4000  \\ [1ex]
      \hline
       $n_{\text{cfg}}\times n_{\text{src}}$ & $504\times16$ &  $454\times8$& $118\times 4$  \\ [1ex]
    
      \hline
      $T/a$ & 14 &  20 & 33 \\ [1ex]
      &17 & 25&40 \\ [1ex]
      & 20& 30& \\ [1ex]

      \hline
    \end{tabular}
  \end{center}
  \label{tab:simulation}
\end{table}

\begin{figure}
  \includegraphics[width=0.4\textwidth]{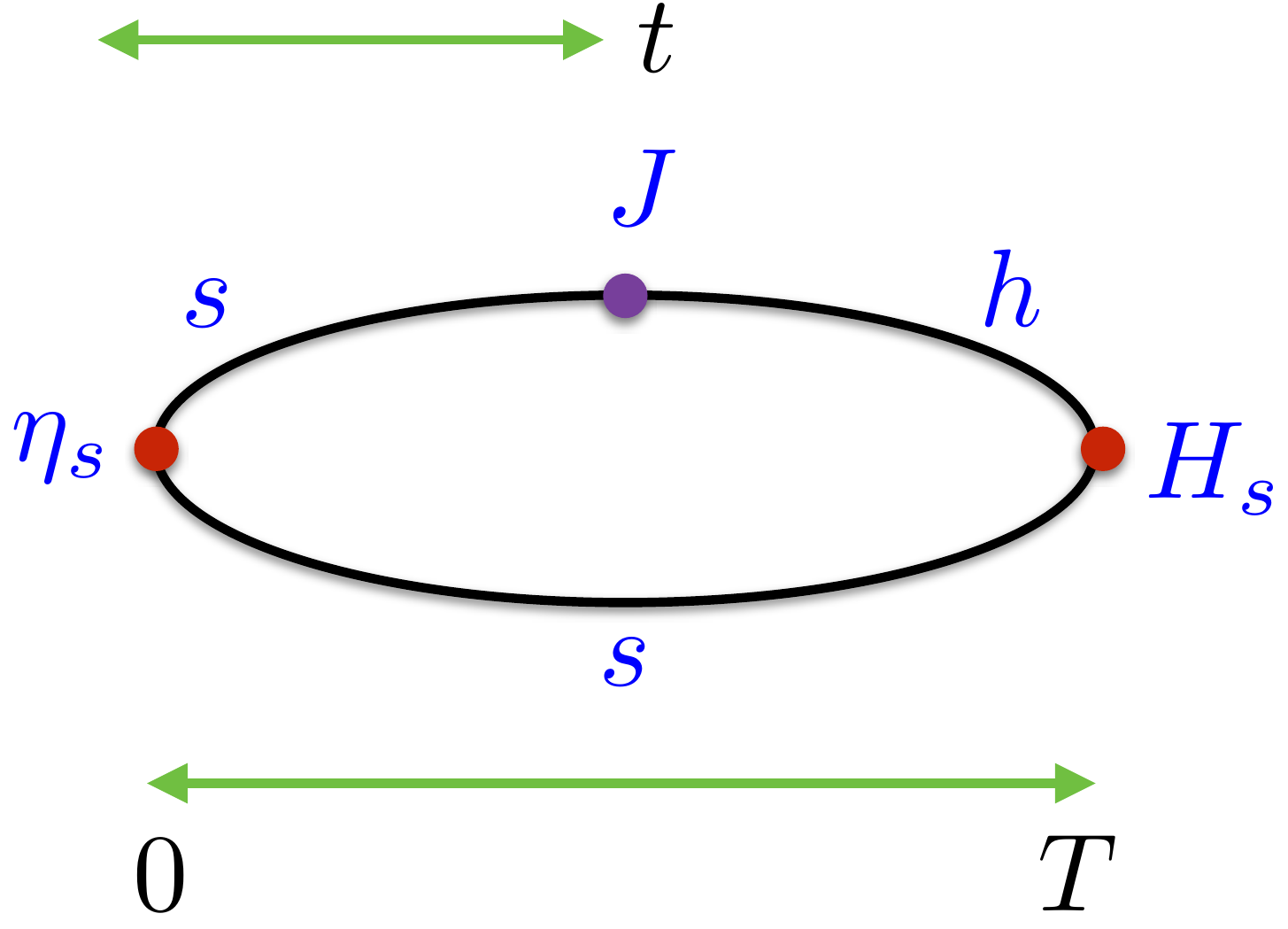}
  \caption{
  The arrangement of propagators in our calculation of the three-point correlation functions.
  }
  \label{Fig:3pt}
\end{figure}

We calculate two-point correlation functions for the Goldstone pseudoscalar ($\gamma^5\otimes\xi^5$) $\eta_s$, and the two heavy-strange bilinears detailed above. 
The correlators are built using
\begin{equation}
  C_{H_s}(t)=\frac{1}{4}\sum_{\vec{x}_0,\vec{x}_t}\langle\text{Tr}[g_h^{\dagger}(x_t,x_0)g_s(x_t,x_0)] \rangle,
\end{equation}
\begin{equation}
  C^{\vec{p}}_{\eta_s}(t)=\frac{1}{4}\sum_{\vec{x}_0,\vec{x}_t}\langle\text{Tr}[g_s^{\theta\dagger}(x_t,x_0)g_s(x_t,x_0)] \rangle,
\end{equation}
where $g_q(x_t,x_0)$ is the one-spinor component staggered propagator for a quark of flavour $q$, from point $x_0 = (0, \vec{x}_0)$ to point $x_t = (t, \vec{x}_t)$. 
The twist angle $\theta$ is given by $\theta = |a\vec{p}\,| N_s / (\sqrt{3} \pi)$, with $a\vec{p}$ in the spatial $(1,1,1)$ direction. 
We sum the spatial components of $x_t$ over the lattice sites to give the two-point correlation function for each $0 \leq t \leq N_t$. 
The $\langle \, \rangle$ denotes path integration over all fields, carried out using the averaging over ensembles, and the trace is over colour. 
Random wall sources are used at $x_0$ to improve statistical precision. 

The local non-Goldstone pseudoscalar ($\gamma^5 \gamma^0 \otimes \xi^5 \xi^0$) heavy-strange meson is similarly defined, but the spin-taste structure is implemented using a lattice site-dependent phase,
\begin{equation}
  C_{\widehat{H}_s}(t) = \frac{1}{4}\sum_{\vec{x}_0, \vec{x}_t} \langle (-1)^{\bar{x}_0^0 + \bar{x}_t^0} \text{Tr} [ g_h^{\dagger} (x_t, x_0) g_s(x_t, x_0)] \rangle,
\end{equation}
where $\bar{x}^{\mu} = (\sum_{\nu\neq\mu} x^{\nu}) / a$. 
We need to use this in the three-point correlation function with temporal vector current in order to cancel tastes. 
The mass of the local non-Goldstone meson only differs from that of the Goldstone by discretisation effects which are very small, and disappear in the limit of zero lattice spacing. 
In our case the mass splittings between $H_s$ and $\widehat{H}_s$ are so small as to only be visible above the statistical uncertainty on the fine lattice.  

We also calculate three-point functions, with the scalar and temporal vector current insertions as defined in Sec.~\ref{Sec:formfactors}. 
We place the $\eta_s$ operator at $x_0$, the current at $x_t$, and the relevant heavy-strange $H_s$ or $\widehat{H}_s$ at $x_T = (T, \vec{x}_T)$, where we again sum over spatial components. 
We then need extended heavy quark propagators from $x_T$ to $x_t$ for each heavy quark mass. 
The three-point functions combine quark propagators as: 
\begin{align}
  C^{\vec{p}}_{S}(t, T) &= \frac{1}{4}\sum_{\vec{x}_0,\vec{x}_t,\vec{x}_T}\langle\text{Tr}[g_h^{\dagger}(x_T,x_t)g_s(x_T,x_0)g^{\theta\dagger}_s(x_t,x_0)] \rangle, \\
    C^{\vec{p}}_{V^0}(t, T) &= \frac{1}{4}\sum_{\vec{x}_0,\vec{x}_t,\vec{x}_T}\langle(-1)^{\bar{x}_t^0+\bar{x}_T^0} \nonumber \\
    &\times \text{Tr} [ g_h^{\dagger}(x_T, x_t) g_s(x_T, x_0) g^{\theta\dagger}_s(x_t, x_0)] \rangle.
\end{align}
$T$ takes several different values on each lattice, detailed in Table~\ref{tab:simulation}, and we determine correlation functions for all $x_t$ from 0 to $T$. 
The combination of propagators needed is illustrated in Fig.~\ref{Fig:3pt}.

\subsection{Analysis of correlation functions}\label{sec:analysis}
%
\begin{figure}
\hspace{-5pt}
  \includegraphics[width=0.49\textwidth]{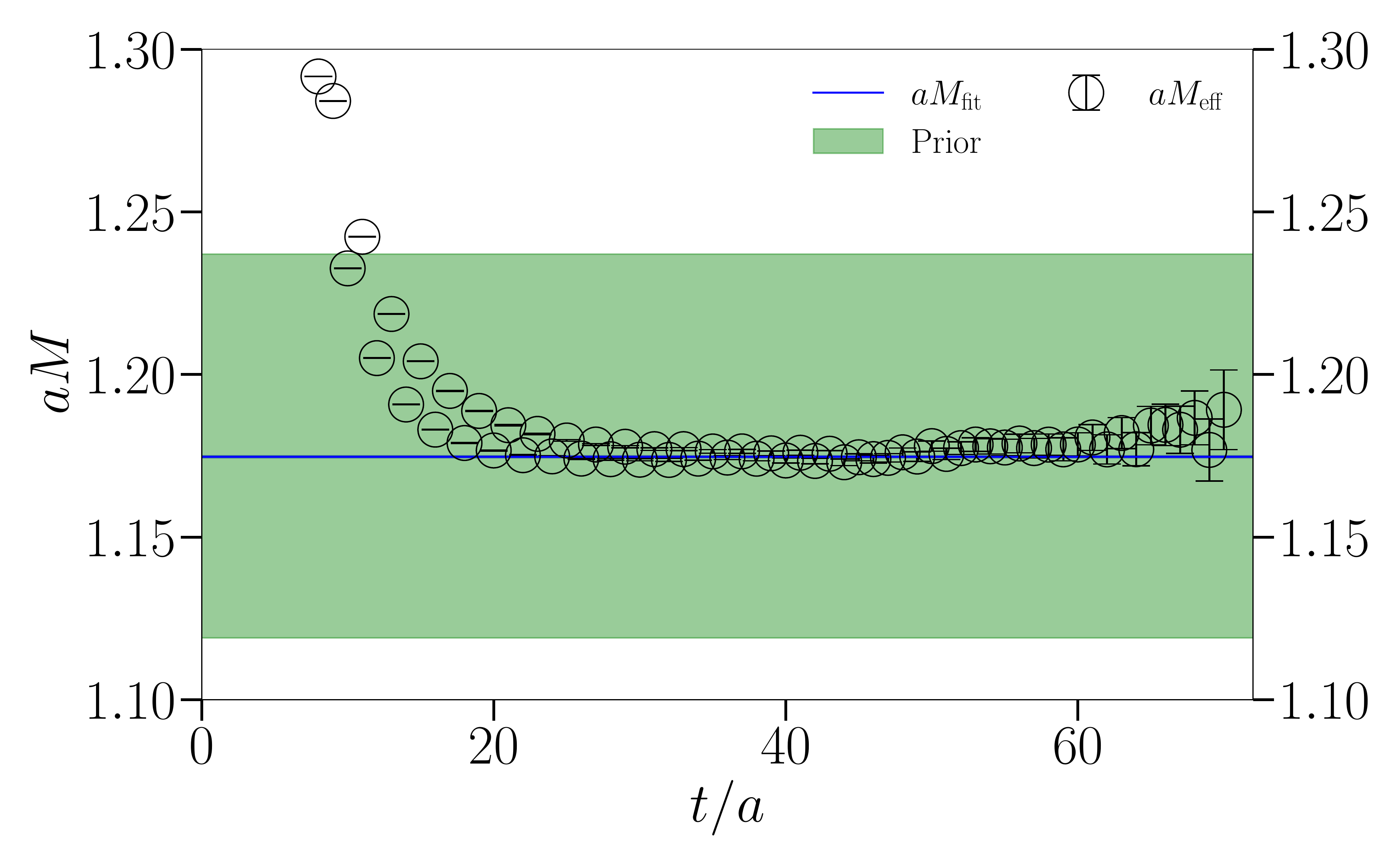}
\hspace{-10pt}
  \includegraphics[width=0.49\textwidth]{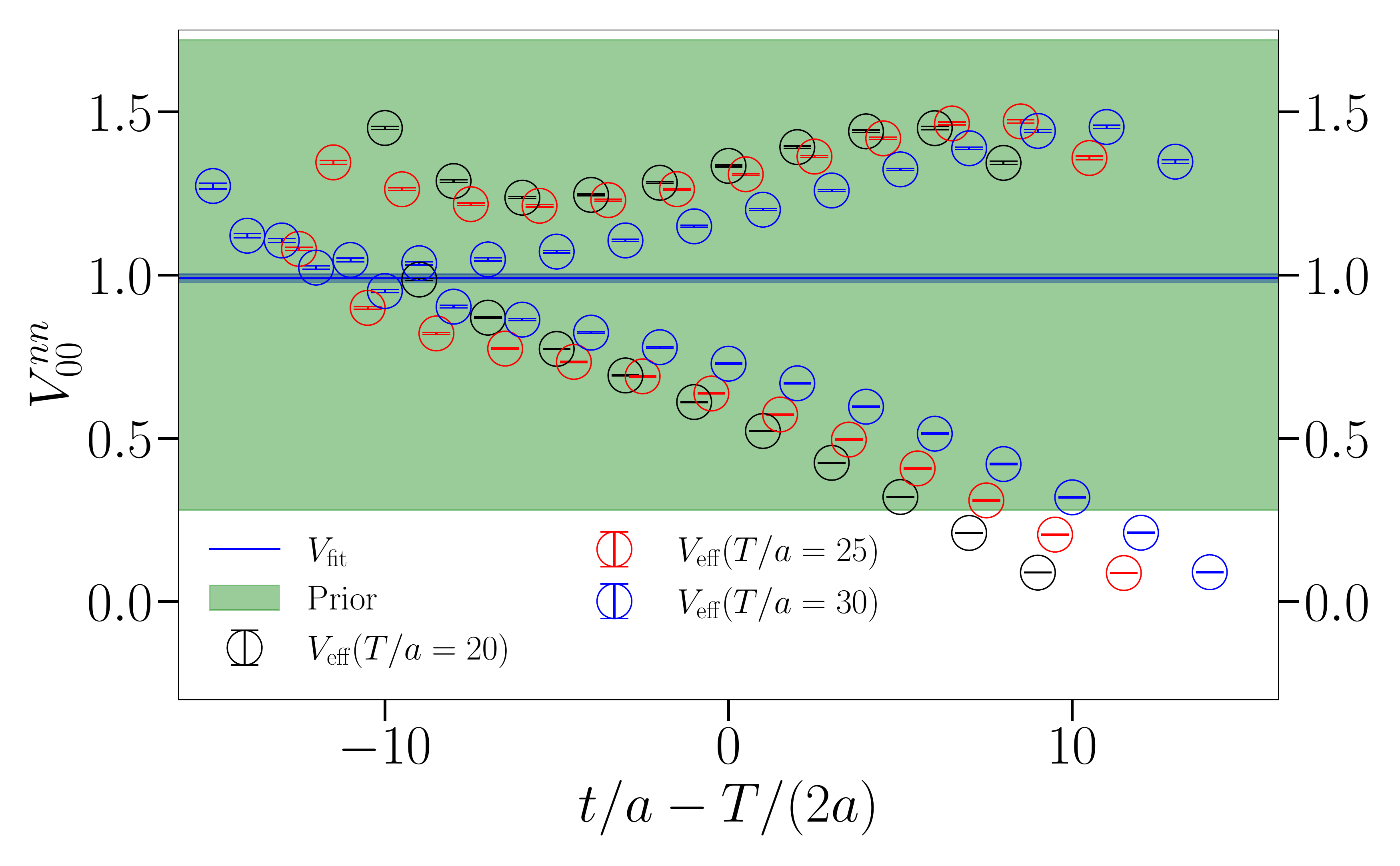}
  \caption{
  Representative plots demonstrating two-point and three-point correlator data, prior selection and fit results.
  Both plots are from set 2 with $am_h = 0.8$.
  In the top panel, Eq.~(\ref{Eq:Meff}) is used to plot the effective mass for the $H_s$ meson two-point correlator data.
  The $E_0^{H_s, n}$ prior is shown by the wide green band and the posterior by the narrow blue band.
  The bottom panel shows the vector three-point correlator data, for $|a{\vec{p}}_{\eta_s}| = 0.143$, plotted as the three-point effective amplitude using Eq.~(\ref{eq:Jeff}).
  The prior for $V_{00}^{nn}$ is given by the wide green band and the posterior by the narrow blue band.}
  \label{Fig:fits}
\end{figure}
We perform a simultaneous, multiexponential fit of the two- and three-point correlation function data using a standard Bayesian approach, introduced in~\cite{Lepage:2001ym} and expanded upon in \cite{PhysRevD.85.031504, Bouchard:2014ypa}.
Further detail is available in the documentation for the Gvar~\cite{peter_lepage_2020_3715065}, Lsqfit~\cite{peter_lepage_2020_3707868} and Corrfitter~\cite{peter_lepage_2019_3563090} Python packages used to perform the analysis.

\begin{figure}
  \hspace{-10pt}
  \includegraphics[width=0.49\textwidth]{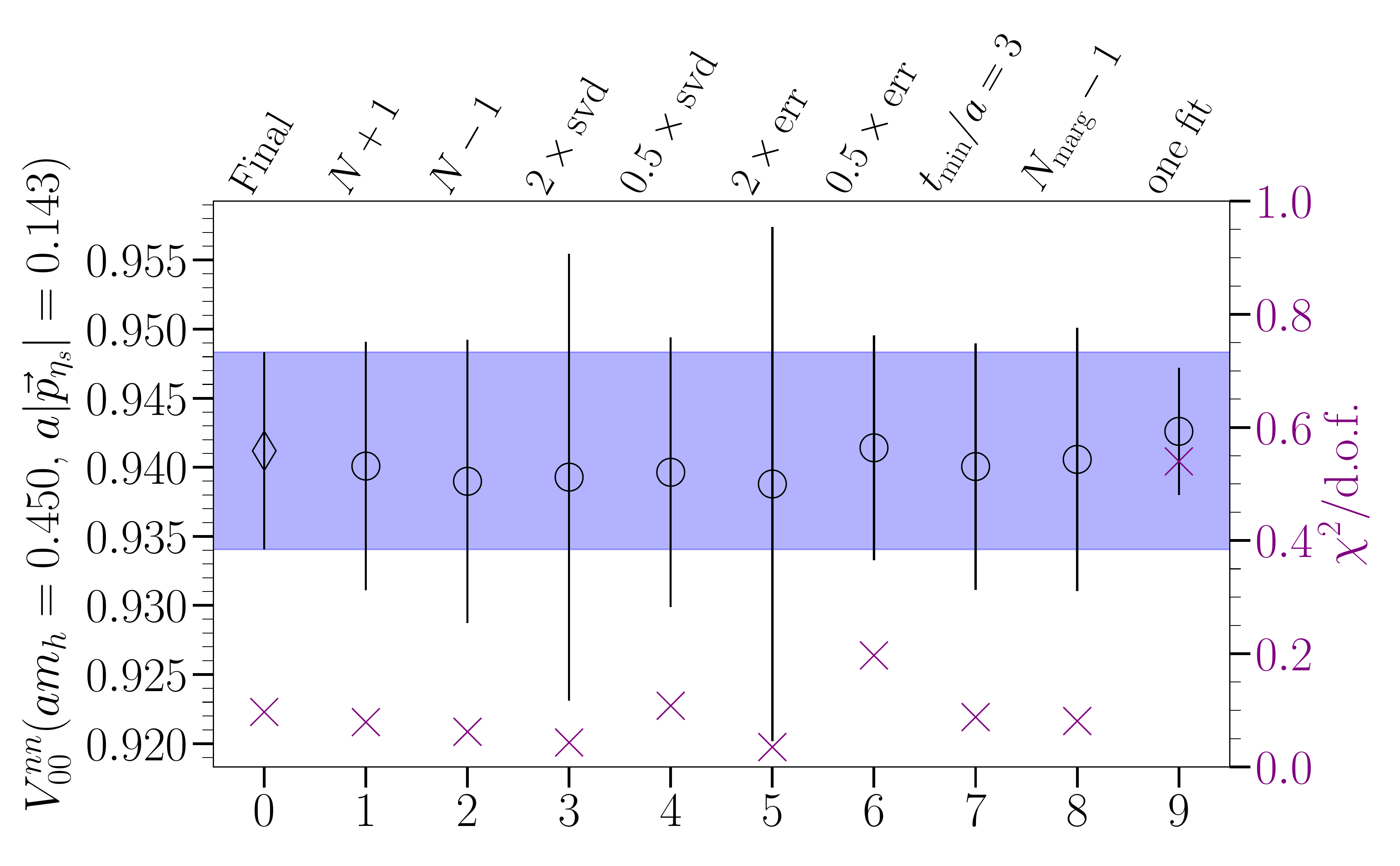}
  \caption{
  Stability tests of the chained, marginalised fit used on a typical three-point correlator. 
  Test 0, the final result, shows the value of $V^{nn}_{00}$ for $am_h = 0.45$, $a|\vec{p}| = 0.1430$ on set 2, with $N^{3\text{pts}}_{\text{exp}} = 3$ exponential terms and three additional states marginalised (as discussed in the text),
  with $t_{\rm min}/a = 2$, the number of data points removed from the fit at the start and end of the data. 
  Tests 1 and 2 show the effects of increasing and decreasing the number of fitted exponentials by 1, tests 3 and 4 show the effect of doubling and halving the SVD cut, and 5 and 6 show the effect of doubling and halving the error on all priors. 
  Test 7 shows the effect of an increase on $t_{\rm min}/a$ by 1, and test 8 shows the reduction of the marginalised exponentials from 6 to 5. 
  Finally, test 9 shows the result of just fitting the vector 3 point correlator for this mass and twist, and the relevant 2 points; this gives a reduced error since the smaller fit requires a smaller SVD cut. 
  Fitting like this does not preserve correlations, however, so we use a global fit. 
  Other two and three-point correlators behaved similarly well under the same tests. 
  The $\chi^2$/dof values (purple $\times$s) are also plotted for reference.
 Note that these are the raw values and hence artificially small (see text) and the degrees of freedom are not the same across all tests.
}
  \label{Fig:fitstability}
\end{figure}

Bias in the small eigenvalues of a large data covariance matrix with a finite data sample is addressed with a singular value decomposition (SVD) cut.
This is a conservative move which avoids underestimating errors (see Appendix D of~\cite{Dowdall:2019bea}). 
We implement the SVD cut by replacing eigenvalues smaller than the product of the cut and the largest eigenvalue with that value. 
We determine an appropriate SVD cut from eigenvalues of bootstrapped data, a facility which is built into Corrfitter. 
We check stability against doubling and halving the SVD cut compared to the recommended value and demonstrate this in Fig.~\ref{Fig:fitstability}.

Using an SVD cut and broad priors can lead to an artificial reduction in $\chi^2$/d.o.f. 
Corrfitter has a built-in facility permitting the determination of a more realistic value (see documentation~\cite{peter_lepage_2020_3707868, peter_lepage_2020_3715065, peter_lepage_2019_3563090} and Appendix D of~\cite{Dowdall:2019bea}) by adding SVD and prior noise. 
We have checked that the fits reported below give values of $\chi^2$/d.o.f. close to 1 with this augmented noise. 
We report the raw $\chi^2$/d.o.f. values in Fig.~\ref{Fig:fitstability} since they still provide a useful comparison between fits.

Bayesian fits provide an additional fit statistic, the Bayes factor, which
penalizes overfitting, thereby providing a measure of fit quality complementary to $\chi^2$.
For each fit, Corrfitter calculates the Gaussian Bayes factor (GBF), the Bayes factor under assumed Gaussian probability distributions.
When evaluated together, GBF and $\chi^2/$d.o.f. provide a useful diagnostic for evaluating the ability of a fit to describe the data while not overfitting.

We aim to extract the ground state energies from the two-point functions, and use these, combined with ground state amplitudes, to extract ground state to ground state matrix elements from the three-point correlators.

We fit two-point correlators for a meson $M$ to
\begin{align}\label{Eq:2ptcorrfitform}
  C^M_{2}(t)&=\sum_{i=0}^{N^{\rm 2pt}_{\text{exp}}}\Big(|a_i^{M,n}|^2\, (e^{-E^{M,n}_i t} + e^{-E^{M,n}_i (N_t-t)}) \nonumber \\
  &-(-1)^t|a_i^{M,o}|^2\, (e^{-E^{M,o}_i t}+e^{-E^{M,o}_i (N_t-t)}) \Big) ,
\end{align}
where a tower of excited states of energy $E^{M,n}_i$ and amplitude $a_i^{M,n}$ above the ground state ($i = 0$) are generated by our lattice operator. 
Discarding data for $t < t_{\rm min}$ allows us to fit a finite number $N^{\rm 2pt}_{\rm exp}$ of these states, and $t_{\rm min}/a$ takes values in the range 3--9 for different correlators and different lattice spacings. 
As detailed in~\cite{2007PhRvD..75e4502F}, HISQ two-point correlators also produce states which oscillate in time from lattice site to lattice site, with the exception of the zero momentum $\eta_s$, where the quark and antiquark of the same mass prevent this effect from being exhibited. 
These states have their own amplitudes and energies $a_i^{M,o}$ and $E^{M,o}_i$ in our fits.

We determine priors for the ground state energies and amplitudes using the effective mass and effective amplitude, defined as
\begin{align}
aM_{\text{eff}}(t) &= \frac{1}{2}\cosh^{-1}\Bigg(\frac{C_2(t-2)+C_2(t+2)}{2C_2(t)}\Bigg),
\label{Eq:Meff} \\
a_{\text{eff}}(t) & = \sqrt{\frac{C_2(t)}{e^{-M_{\text{eff}}t}+e^{-M_{\text{eff}}(N_t-t)}}}.
\label{Eq:Aeff}
\end{align}
We give each a broad uncertainty, checking that the final result of the fit is much more precisely determined than this prior. 
The lowest oscillating state prior is taken to be the nonoscillating ground state plus $\Lambda_{\rm QCD}$ (which we take to be 0.5\,GeV), with an error around 50\%. 
The energy differences between all excited states are taken to be $\Lambda_{\rm QCD}$ with an error of 50\%. 
We use log-normal priors throughout to enforce positive values on energy splittings and amplitudes. 
Priors for excited state nonoscillating and all oscillating amplitudes are based on previous experience of amplitude sizes, and some are slightly adjusted to maximise the GBF; these are listed in Table~\ref{Tab:priorsforfit}. 
In all cases, priors are many times broader than the final fit uncertainties, as demonstrated in Fig.~\ref{Fig:fits}. 

We perform three-point fits to
\begin{equation}\label{Eq:3ptcorrfitform}
\begin{split}
  &C_3(t,T) = \sum_{i,j=0}^{N^{\rm 3pt}_{\text{exp}}} \Big( a_i^{\eta_s,n} J_{ij}^{nn} a_j^{\brachat{H_s},n}\, e^{-E^{\eta_s,n}_i t}\, e^{-E^{\brachat{H_s},n}_i (T-t)}\\
  &-(-1)^{T-t}\, a_i^{\eta_s,n}J_{ij}^{no}a_j^{\brachat{H_s},o}\, e^{-E^{\eta_s,n}_i t}\, e^{-E^{\brachat{H_s},o}_i (T-t)}\\
  &-(-1)^{t}\, a_i^{\eta_s,o} J_{ij}^{on} a_j^{\brachat{H_s},n}\, e^{-E^{\eta_s,o}_it}\, e^{-E^{\brachat{H_s},n}_i (T-t)}\\
  &+(-1)^{T}\, a_i^{\eta_s,o} J_{ij}^{oo} a_j^{\brachat{H_s},o}\, e^{-E^{\eta_s,o}_i t}\, e^{-E^{\brachat{H_s},o}_i (T-t)}\Big) ,
\end{split}
\end{equation}
for different masses of $H_s$ (for the scalar current insertion) or $\widehat{H}_s$ (for the temporal vector current insertion) mesons and different twists of $\eta_s$ meson. 
$J_{ij}^{no}$ represents the amplitude for the $i$th nonoscillating state of the $\eta_s$ and the $j$th oscillating state of the heavy meson. 
$J = S, V$, for our scalar and vector current insertions. 
We create the $\eta_s$ at $t = 0$, insert the current at $t$ and annihilate the $H_s$ ($\widehat{H}_s$) at $T$.

Priors for $J_{00}^{nn}$ are based on the effective three-point amplitudes, which can be determined from
\begin{align}
J_{\rm eff}(t, T) = \frac{C_3(t, T)}{a_{\rm eff}^{\eta_s}\, a_{\rm eff}^{\brachat{H_s}}} \, e^{M_{\rm eff}^{\eta_s} t}\, e^{M_{\rm eff}^{\brachat{H_s}} (T-t)}.
\label{eq:Jeff}
\end{align}
Priors for all other $J_{ij}^{kl}$ values are listed in Table~\ref{Tab:priorsforfit}. 
Fig.~\ref{Fig:fits} shows representative plots of the two-point and three-point correlator data, illustrating prior selection and providing a comparison of fit results with both the prior and data.
The effect of doubling and halving the standard deviation given to all priors on the overall results of the fit are shown in Fig.~\ref{Fig:fitstability}.

\begin{table}[t]
  \caption{
  Priors used in the fit on each set. 
  Priors are based on previous experience and given large widths. 
  In some places, adjustment is made for lattice spacing, and priors are tuned using an increase in the GBF. 
  The effect of doubling and halving the standard deviation on all priors on the final fit result is shown in Fig.~\ref{Fig:fitstability}.
  }
  \begin{center} 
    \begin{tabular}{c c c c c c c c c c }
      \hline
      Set & $a^{n}_{i\neq{}0} \text{\ and\ } a^o_i$  & $S_{00}^{kl\neq{}nn}$ & $V_{00}^{kl\neq{}nn}$& $S^{kl}_{ij\neq{}00}$ & $V^{kl}_{ij\neq{}00}$   \\ [0.5ex]
      \hline
      1 & 0.10(10) &0.0(8)&0.2(1.0)&0.0(3)&0.0(3) \\ [1ex]
      2 & 0.10(10) &0.0(8)&0.0(1.0)&0.0(3)&0.0(4) \\ [1ex]
      3 & 0.05(05) &0.0(8)&0.0(1.0)&0.0(3)&0.0(4) \\ [1ex]
      \hline
    \end{tabular}
  \end{center}
  \label{Tab:priorsforfit}
\end{table}

On each ensemble, we perform a chained, marginalised fit to our two- and three-point correlator data.  
For detailed descriptions of chaining and marginalisation, see~\cite{PhysRevD.85.031504, Bouchard:2014ypa} and the Corrfitter documentation~\cite{peter_lepage_2019_3563090}.

The chained fit begins with a simultaneous fit to all of the two-point correlators ($H_s$ and $\widehat{H}_s$ for each $m_h$ and $\eta_s$ for each $a\vec{p}$), fixing $N^{\rm 2pt}_{\text{exp}}$ in Eq.~(\ref{Eq:2ptcorrfitform}) for each lattice spacing such that it gives an acceptable $\chi^2$ and maximises the GBF.  
We take $N^{\rm 2pt}_{\rm exp} = 5$ in the case of set 1 and $N^{\rm 2pt}_{\rm exp} = 6$ in the case of sets 2 and 3. 
The next step in the chained fit is a simultaneous fit to all three-point correlators.
This includes both $S$ and $V$ current insertions and data at the values for $T$ chosen for each ensemble (listed in Table~\ref{tab:simulation}).
The chained fit prescription uses two-point correlator fit posteriors as priors for the two-point fit parameters that appear in the subsequent three-point correlator fit, accounting for correlations between these posteriors and the three-point correlator data.

In the three-point correlator fits, the number of states $N^{\rm 3pt}_{\rm exp}$ in Eq.~(\ref{Eq:3ptcorrfitform}) must be understood in terms of marginalisation.
Marginalisation~\cite{PhysRevD.85.031504} subtracts fit function terms, evaluated using priors, from the data before performing the fit.
In this way, effects from these terms are accounted for while the fit function used by the minimiser is simplified.
For sets 1, 2 and 3, we choose $N^{\rm 3pt}_{\rm exp} = 2$, 3 and 2, respectively, such that we achieve an acceptable fit ($\chi^2$ per degree of freedom of 0.342, 0.079 and 0.047, respectively.)
On each set, the total number of states accounted for, either explicitly fit using Eq.~(\ref{Eq:3ptcorrfitform}) or subtracted from the data, is equal to $N^{\rm 2pt}_{\rm exp}$.
For example, on set 1 we fit two-point correlators with $N^{\rm 2pt}_{\rm exp} = 5$.
For the fit to the three-point correlators, we first subtract from the data contributions from terms in Eq.~(\ref{Eq:3ptcorrfitform}) with $i$ or $j$ equal to 3, 4 or 5.
We then fit this data using Eq.~(\ref{Eq:3ptcorrfitform}) with $N^{\rm 3pt}_{\rm exp} = 2$.
This is useful here because our three-point data are noisier than our two-point data, so fewer states are required in their fits.
Marginalisation allows us to include information about higher states obtained from two-point fits.

\begin{figure}
  \hspace{-10pt}
  \includegraphics[width=0.49\textwidth]{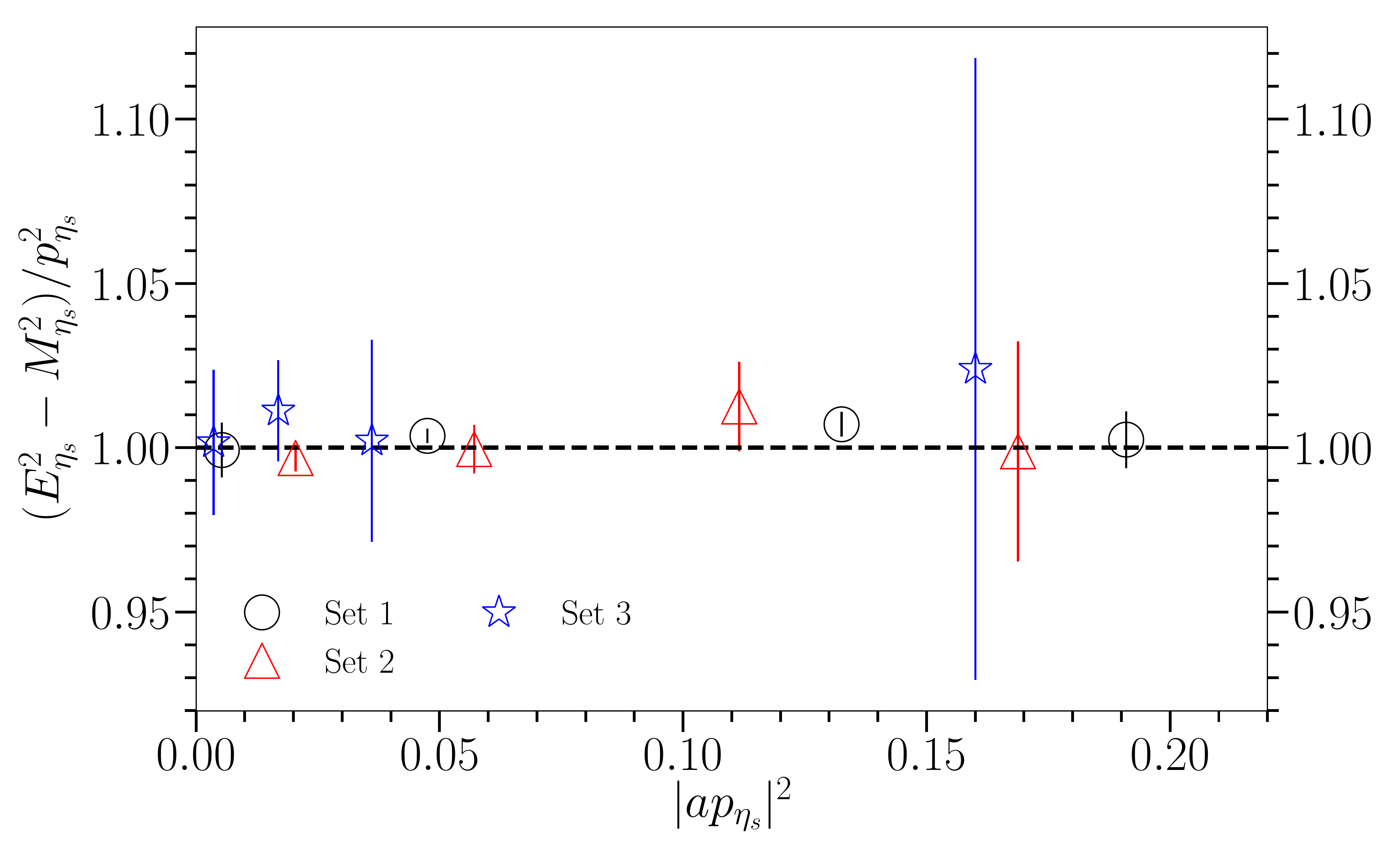}
  \caption{
We plot the ratio $(E_{\eta_s}^2 - M_{\eta_s}^2) / |\vec{p}_{\eta_s}|^2$ from our fit results against $|a\vec{p}_{\eta_s}|^2$ to check that the $\eta_s$ energy in our final fit results agrees with the momentum given to the meson in the lattice calculation. 
The results agree well throughout the range of momenta.}
  \label{Fig:speedoflight}
\end{figure}

We also check that the momentum dispersion relation for our $\eta_s$ fit results agrees with the momenta given in the lattice calculation. 
The two should differ by discretisation effects only, which are small for the $\eta_s$ as it contains only $s$ quarks but grow with momentum. 
This is displayed in Fig.~\ref{Fig:speedoflight}.

Fit results are converted according to
\begin{equation}\label{Eq:fitnormalisation}
  \bra{\eta_s} J \ket{\brachat{H_s}} = 2\sqrt{M_{H_s} E_{\eta_s}} J^{nn}_{00},
\end{equation}
to extract the matrix elements which appear in the definition of the form factors [Eqs.~(\ref{Eq:vec}) and~(\ref{Eq:sca})]. 
We always use the mass of the Goldstone heavy-strange pseudoscalar for $M_{H_s}$ as the non-Goldstone mass is the same in the continuum limit. 
Tests showed that changing this to the non-Goldstone mass in the case of the vector matrix element made no difference at all to our continuum form factors, as discretisation errors are accounted for in our extrapolation to the physical point.  

The results from the fits for each of the three lattice spacings are summarised in the Appendix in Tables~\ref{tab:finefitresults}--\ref{tab:ultrafinefitresults}.

\subsection{Current normalisation}\label{sec:norm}
The PCVC relation, applied at zero spatial momentum for the daughter meson,
\begin{equation}\label{Eq:Zv}
    Z_V^{0} = \frac{(m_h - m_s) \bra{\eta_s} S \ket{H_s}}{(M_{H_s} - M_{\eta_s})\bra{\eta_s} V^{0} \ket{H_s}} \Bigg\rvert_{\vec{p}_{\eta_s} = 0},
\end{equation}
allows us to normalise the vector matrix element nonperturbatively using the scalar matrix element~\cite{Koponen:2013tua, Donald:2013pea, McLean:2019qcx}. 
This uses the fact that this current is conserved in the HISQ formalism, that is to say that the product of the bare mass and the scalar matrix element does not require renormalisation.
We also make the small correction $Z_{\text{disc}}$ to account for small tree-level mass-dependent discretisation effects beginning at order $(am_h)^4$. 
For the determination of $Z_{\text{disc}}$ see~\cite{Monahan:2012dq,McLean:2019sds}. 
Values for these normalisations can be found in Table~\ref{tab:renorm}. 
\begin{table}[t]
  \caption{
  Values for normalisation constants appearing in Eqs.~(\ref{Eq:vec}) and~(\ref{Eq:sca}). 
  $Z_V$ is calculated as in Eq.~(\ref{Eq:Zv}) and $Z_{\text{disc}}$ is defined in~\cite{Monahan:2012dq}.
  }
  \begin{center} 
    \begin{tabular}{c c c c}
      \hline
      Set &  $am^{\text{val}}_{h}$& $Z_V$ &$Z_{\text{disc}}$\\ [0.5ex]
      \hline
      1 & 0.449 &1.0061(25)& 0.99892\\ [1ex]
        & 0.566 &1.0110(30)&  0.99826  \\ [1ex]
        & 0.683 &1.0164(36)& 0.99648  \\ [1ex]
        & 0.8 &1.0226(43)& 0.99377   \\ [1ex]   

      \hline
      2 & 0.274 &1.0003(73)&0.99990\\ [1ex]
        & 0.45 &1.004(10)& 0.99928 \\ [1ex]
        & 0.6 &1.008(12)& 0.99783  \\ [1ex]
        & 0.8 &1.018(15)& 0.99377  \\ [1ex]

      \hline
      3 &  0.194 &0.996(27)&0.99997\\ [1ex]
       & 0.45 &0.987(55)& 0.99928  \\ [1ex]
       & 0.6 &1.015(73)& 0.99783\\ [1ex]
       & 0.8 &1.032(96)& 0.99377  \\ [1ex]

      \hline
    \end{tabular}
  \end{center}
  \label{tab:renorm}
\end{table}
%

\subsection{Continuum and quark mass extrapolation}\label{sec:extrap}
Having calculated $f_0(q^2)$ and $f_+(q^2)$ for the three lattice spacings and at a range of heavy quark masses and $q^2$ values on each lattice, we now perform a fit in heavy quark mass, sea quark mass and lattice spacing. 
We can then evaluate our form factors at the physical quark masses and zero lattice spacing. 
Our fits also allow us to examine the heavy quark mass dependence of the form factors, all the way down to the charm mass.

\subsubsection{Fit ansatz and priors}\label{sec:ansatz_priors}
Following the method successfully employed in~\cite{McLean:2019qcx}, we fit the form factors on the lattice using the Bourreley-Caprini-Lellouch (BCL) parameterisation~\cite{Bourrely:2008za}:
\begin{equation}\label{Eq:zexpansion}
\begin{split}
  f_0(q^2)&=\frac{1}{1-\frac{q^2}{M^2_{H_{s}^{0}}}}\sum_{n=0}^{N-1}a_n^0z^n,\\
  f_+(q^2)&=\frac{1}{1-\frac{q^2}{M^2_{H_{s}^{*}}}}\sum_{n=0}^{N-1}a_n^+\Big(z^n-\frac{n}{N}(-1)^{n-N}z^N\Big),
\end{split}
\end{equation}
where we use a mapping of $q^2$ to $z$, a region inside the unit circle of the $z$ plane,
\begin{equation}
  z(q^2)=\frac{\sqrt{t_+-q^2}-\sqrt{t_+-t_0}}{\sqrt{t_+-q^2}+\sqrt{t_+-t_0}},
\end{equation}
with $t_+ = (M_{H} + M_{K})^2$, the lowest mass combination with the same quantum numbers as the current where a cut in the $q^2$ plane begins. 
Since we do not explicitly determine $M_H$ and $M_K$ here, we use $M_H = M_{H_s} + a(M_B^{\rm phys} - M_{B_s}^{\rm phys})$ and $M_K = M_{\eta_s} + a(M_K^{\rm phys} - M_{\eta_s}^{\rm phys})$. 
These have the correct limit at physical quark mass values. 
We choose to take $t_0 = 0$.
To fit the data for $f_0(q^2)$ and $f_+(q^2)$, tabulated in the Appendix, we calculate for each quark mass and momentum simulated the corresponding value of $z$, using the associated meson masses and values of $q^2$.

The poles in Eq.~(\ref{Eq:zexpansion}) account for the production of on shell $H_{s0}$ and $H_s^*$ states for $q^2 > q^2_{\rm max}$, and the mass $M_{H_{s0}}$ is taken as $M_{H_s} + 0.4\,\text{GeV}$, which is consistent with lattice results in~\cite{Dowdall:2012ab} and experimental results~\cite{PDG} for the axial vector--vector splitting, $M_{B_s}(1^+) - M_{B_s}(1^-)$. 
We do not need to know this number precisely as we are simply removing a pole in the data to ease the fitting process and then replacing it later. 
Indeed, excluding the pole from the $f_0$ fit function completely leads to fit results which are consistent with those from including the pole. 
The position of the $M_{H_s^*}$ pole can be estimated, as in~\cite{McLean:2019qcx}, using the fact that $M_{H_s^*} - M_{H_s} \to 0$ as $m_h \to \infty$, with the ansatz $M_{H_s^*} = M_{H_s} + x / M_{H_s}$. 
We find $x$ from the Particle Data Group (PDG~\cite{PDG}) value of $M^{\rm phys}_{B_s^*} - M^{\rm phys}_{B_s} = x/M^{\rm phys}_{B_s} = 0.0489(15)$\,GeV. 
We go one step further to ensure that this ansatz also gives the correct PDG value for $M^{\rm phys}_{D^*_s} = 2.1122(4)$\,GeV, using
\begin{align}
    M_{H^*_s} &= M_{H_s}+\frac{M^{\rm phys}_{D_s}}{M_{H_s}}\Delta(D_s) \\
    &+ \frac{M^{\rm phys}_{B_s}}{M_{H_s}} \Big[ \frac{M_{H_s} - M^{\rm phys}_{D_s}}{M^{\rm phys}_{B_s} - M^{\rm phys}_{D_s}} \Big( \Delta(B_s) - \frac{M^{\rm phys}_{D_s}}{M^{\rm phys}_{B_s}} \Delta(D_s) \Big) \Big], \nonumber
\end{align}
with splittings $\Delta(B_s) = 0.0489(15)$\,GeV and $\Delta(D_s) = 0.14386(41)$\,GeV, from the PDG.
We find no significant difference in the final form factors from the change of ansatz, supporting our assertion that the exact pole position is not crucial, as any small errors here are accounted for by higher orders of the $z$ expansion.
We use $N = 3$ in Eq.~(\ref{Eq:zexpansion}) for our final results.   

We fit coefficients $a^{0,+}_n$ to a general fit form, accounting for heavy quark mass dependence and discretisation effects:
\begin{equation}\label{Eq:an}
\begin{split}
  a_n^{0,+} &= \Big(1 + \rho_n^{0,+} \log \Big(\frac{M_{H_s}}{M_{D_s}}\Big) \Big)\times\\
  &\sum^{N_{ijk} - 1}_{i,j,k = 0} d_{ijkn}^{0,+} \Big( \frac{\Lambda_{\rm QCD}}{M_{H_s}} \Big)^i \Big( \frac{am_h^{\rm val}}\pi \Big)^{2j} \Big( \frac{a\Lambda_{\rm QCD}}\pi \Big)^{2k}\\
  &\times(1 + \mathcal{N}^{0,+}_n).
\end{split}
\end{equation}
We use $M_{H_s}$ as a physical proxy for the heavy quark mass, as the two are equal at leading order in HQET. 
Terms in $\Lambda_{\rm QCD} / M_{H_s}$ (with $\Lambda_{\rm QCD} = 0.5$\,GeV) parameterise the effect of changing heavy mass, whilst the other terms in the sum allow for discretisation effects, which for the HISQ action appear as even powers of energy scales. 
In this case the two relevant energies are the heavy quark mass and $\Lambda_{\rm QCD}$.
The log term comes from the matching of our HQET-inspired fit function to QCD~\cite{Banerjee:2017zdm,Bahr:2016ayy}. 
From~\cite{Banerjee:2017zdm}, we expect the coefficient of the log term to be of order unity, so we use a prior of $0\pm1$. 

The term 
\begin{equation}
\begin{split}
  \mathcal{N}_n^{0,+} &= \frac{c_{s,n}^{\text{val},0,+} \delta_s^{\rm val} + c_{s,n}^{0,+} \delta_s + 2c_{l,n}^{0,+} \delta_l}{10m_s^{\rm tuned}}\\
  &+ c_{c,n}^{\text{val},0,+} \Big( \frac{M_{\eta_c} - M_{\eta_c}^{\rm phys}}{M_{\eta_c}^{\rm phys}} \Big),
\end{split}
\label{Eq:Nn}
\end{equation}
accounts for mistuning of valence (marked val) and sea quark masses, where $\delta_q^{({\rm val})}=m_q^{({\rm val})} - m_q^{\rm tuned}$. 
We determine the tuned mass of the strange quark using  
\begin{equation}
  m_s^{\rm tuned} = m_s^{\rm val} \Bigg( \frac{M_{\eta_s}^{\rm phys}}{M_{\eta_s}} \Bigg)^2,
\end{equation}
where $M_{\eta_s}^{\rm phys}= 0.6885(22)$\,GeV was calculated in~\cite{Dowdall:2013rya}. 
We find $m_l^{\rm tuned}$ using~\cite{Bazavov:2017lyh}
\begin{equation}
  \frac{m_s^{\rm phys}}{m_l^{\rm phys}} = 27.18(10).
\end{equation}
We find $M_{\eta_c}$ on the three sets from~\cite{McLean:2019qcx} and take $M_{\eta_c}^{\rm phys} = 2.9766(12)$\,GeV. 
This value differs from the experimental $\eta_c$ mass~\cite{PDG} by 7\,MeV to allow for the effect determined in~\cite{Hatton:2020qhk} of missing quark-line disconnected diagrams in the lattice calculation of the $\eta_c$ mass.

We give all $d$ coefficients a prior of $0\pm1$, with the exception of $d_{i10n}$, which multiply terms with $(am_h)^2$ in them. 
Since the HISQ action is improved up to second order in the lattice spacing, we expect these terms to be small, giving them a prior of $0.0\pm0.5$. 
We set $d^{+}_{i000} = d^{0}_{i000}$ and $\rho^+_0 = \rho_0^0$ to enforce $f_0(0) = f_+(0)$ on the fit, in the continuum and in the absence of quark mistuning, although relaxing this constraint still leaves the two values agreeing within errors, giving $f_+(0)/f_0(0) = 0.95(11)$. 
We take $c_s^{\rm val} = 0\pm1$ based on a study of $s$ quark mistuning. 
In the case of maximum mistuning, where $m_s = m_l$ and we have the $B \to \pi$ decay, we can compare our form factors with those from~\cite{Lattice:2015tia}, and find that this gives an upper bound on our valence quark mistuning of $c_s^{\rm val} \approx 2$. 
This is a very extreme case of quark mistuning, so we take the prior width at half of this. 
Sea quark mistunings, as well as those of the valence charm quark, make less of a contribution so we give all other $c$ coefficients a prior of $0.0 \pm 0.3$. 
In Eq.~(\ref{Eq:an}) we take $N_{ijk} = 3$.

In our fit we also include a data point corresponding to the $B_s \to \eta_s$ scalar form factor in the continuum, $f_0(q^2_{\rm max}) = 0.811(17)$ from previous work by the HPQCD Collaboration~\cite{Colquhoun:2015mfa}. 
This data point was obtained in a calculation using NRQCD $b$ quarks, working directly at the tuned $b$ quark mass. 
A ratio was constructed to remove the systematic errors from renormalisation of the NRQCD-HISQ current that would otherwise reduce the accuracy of the result. 
For this reason, this point can be included alongside our HISQ data, without introducing additional errors associated with NRQCD. 
This result is included as a statistically independent data point for the $f_0$ fit function in the continuum and physical quark mass limits and reduces our error at $f_0(q^2_{\rm max})$.
The effect of its removal is demonstrated by test 4 in Fig.~\ref{Fig:ExtrapStab}.

\subsubsection{Continuum and physical quark mass limit}
The fit outlined in the previous section has a $\chi^2$ value of $0.16$ per degree of freedom, for 109 degrees of freedom.
It produces best-fit results for the coefficients in Eqs.~(\ref{Eq:an}) and~(\ref{Eq:Nn}), from which we construct the $z$-expansion coefficients of Eq.~(\ref{Eq:zexpansion}).

By evaluating Eq.~(\ref{Eq:an}) at $a,\, \mathcal{N}^{0,+}_{n} = 0$, we obtain the $z$-expansion coefficients, and therefore the form factors from Eq.~(\ref{Eq:zexpansion}), in the continuum limit and at physical light, strange and charm quark masses.
By choosing physical values of $M_{B_s}^{\rm phys} = 5.36688(14)$\,GeV, $M_{B^*_s}^{\rm phys} = 5.4158(15)$\,GeV and $M_{D_s}^{\rm phys} = 1.968340(70)$\,GeV from the PDG~\cite{PDG} and $M_{B_{s0}}^{\rm phys} = M_{B_s}^{\rm phys} + 0.4$\,GeV, we ensure the $H_s$ interpolates between the correct physical mass $B_s$ and $D_s$ mesons.

In Table~\ref{tab:ancoefficients} we show the final results of our evaluation of the form factors $f_0^{\rm phys}(q^2)$ and $f_+^{\rm phys}(q^2)$ at the physical point for the $B_s \to \eta_s$ decay. 
From the given values of the coefficients and pole masses, as well as their correlation matrix (given in the bottom of the table), one can fully reconstruct both form factors across the full physical $q^2$ range. 

\subsubsection{Fit analysis and stability check}
\begin{figure*}
  \hspace{-30pt}
  \includegraphics[width=0.96\textwidth]{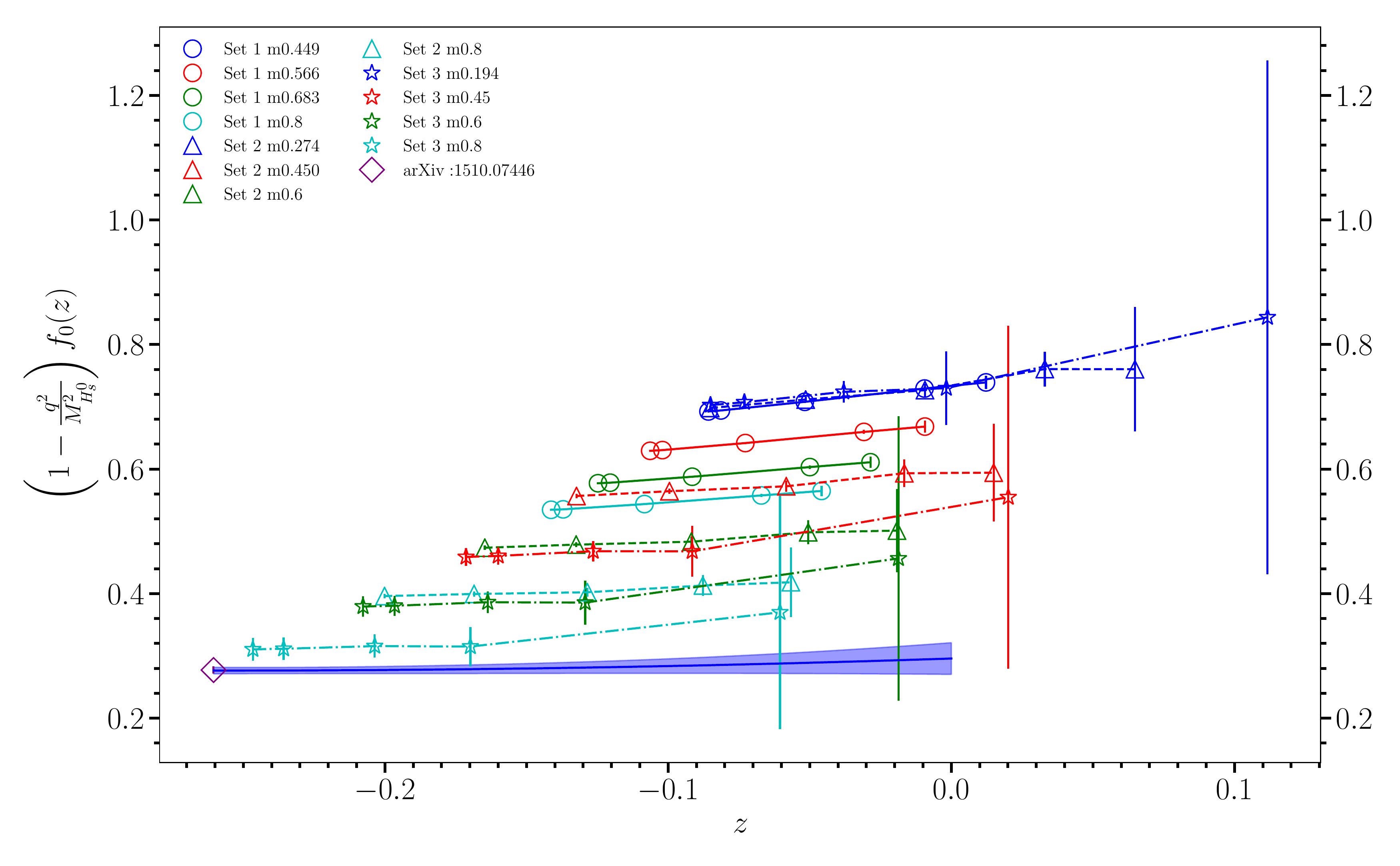}
    \vspace{-0.15in}
  \caption{
  $\Big(1-\frac{q^2}{M^2_{H^0_s}}\Big)f_0(z)$ data points and final result at the physical point (blue band). 
  Data points are labeled by mass for sets 1, 2 and 3, respectively, where e.g. m0.8 indicates $am_h = 0.8$ on that ensemble. 
  Lines between data points of a given heavy mass over the full $z$ range are there to guide the eye. 
  The additional continuum data point from~\cite{Colquhoun:2015mfa} is shown as a purple diamond and helps to pin down the form factor in the high $q^2$ limit.
  }
  \label{Fig:f0nopoleinz}
  \vspace{0.2in}
  \includegraphics[width=0.96\textwidth]{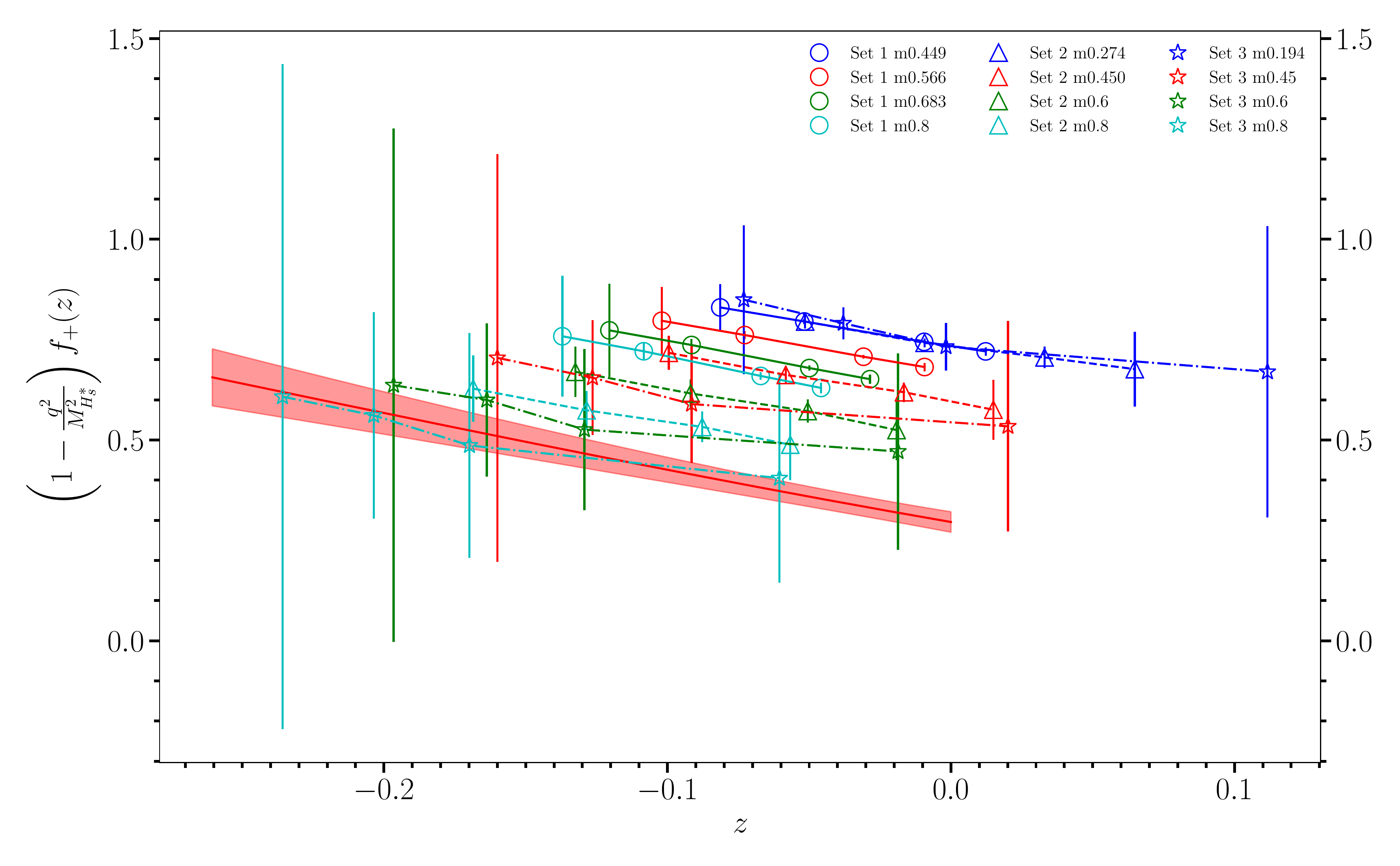}
    \vspace{-0.15in}
  \caption{
  $\Big(1-\frac{q^2}{M^2_{H^*_s}}\Big)f_+(z)$ data points and final result at the physical point (red band). 
  Data points are labeled by mass for sets 1, 2 and 3, respectively, where e.g. m0.8 indicates $am_h = 0.8$ on that ensemble. 
  Lines between data points of a given heavy mass over the full $z$ range are there to guide the eye.
  }
  \label{Fig:fplusnopoleinz}
\end{figure*}
\begin{table*}
  \caption{
  Values of fit coefficients $a_n^{0,+}$ and pole masses at the physical point for the $B_s \to \eta_s$ decay with correlation matrix are given below. 
  Form factors can be reconstructed by evaluating Eq.~(\ref{Eq:zexpansion}) using these coefficients and pole masses.
  Note that $M_{B_{s0}}$ is set to $M_{B_s} + 0.4$\,GeV.
  Masses are in GeV. 
  The pole masses are very slightly correlated due to the way the fit function is constructed. 
  These correlations are too small to have any meaningful effect on the fit, but we include them for completeness in reconstructing our results.
  }
  \begin{center} 
    \begin{tabular}{c c c c c c c c}
      \hline
      $a_0^0$&$a_1^0$&$a_2^0$&$a_0^+$&$a_1^+$&$a_2^+$ & $M_{B_{s0}}$ & $M_{B^*_s}$ \\ [0.5ex]
      0.296(25)&0.15(20)&0.29(47)&0.296(25)&$-1.22(32)$&0.9(1.2)&5.76688(17)&5.4158(15)\\ [1ex]
      \hline 

      1.00000&0.90818&0.72266&1.00000&0.30483&0.09764&$-0.00042$&0.00021\\ [0.5ex]
      &1.00000&0.93763&0.90818&0.38642&0.09064&0.00002&$-0.00009$\\ [0.5ex]
      &&1.00000&0.72266&0.40724&0.07271&0.00012&$-0.00036$\\ [0.5ex]
      &&&1.00000&0.30483&0.09764&$-0.00042$&0.00021\\ [0.5ex]
      &&&&1.00000&0.51317&0.00179&$-0.01229$\\ [0.5ex]
      &&&&&1.00000&$-0.00045$&0.00248\\ [0.5ex]
      &&&&&&1.00000&0.00000\\ [0.5ex]
      &&&&&&&1.00000\\ [0.5ex]
      \hline
    \end{tabular}
  \end{center}
  \label{tab:ancoefficients}
\end{table*}
\begin{figure}
  \hspace{-12pt}
  \includegraphics[width=0.50\textwidth]{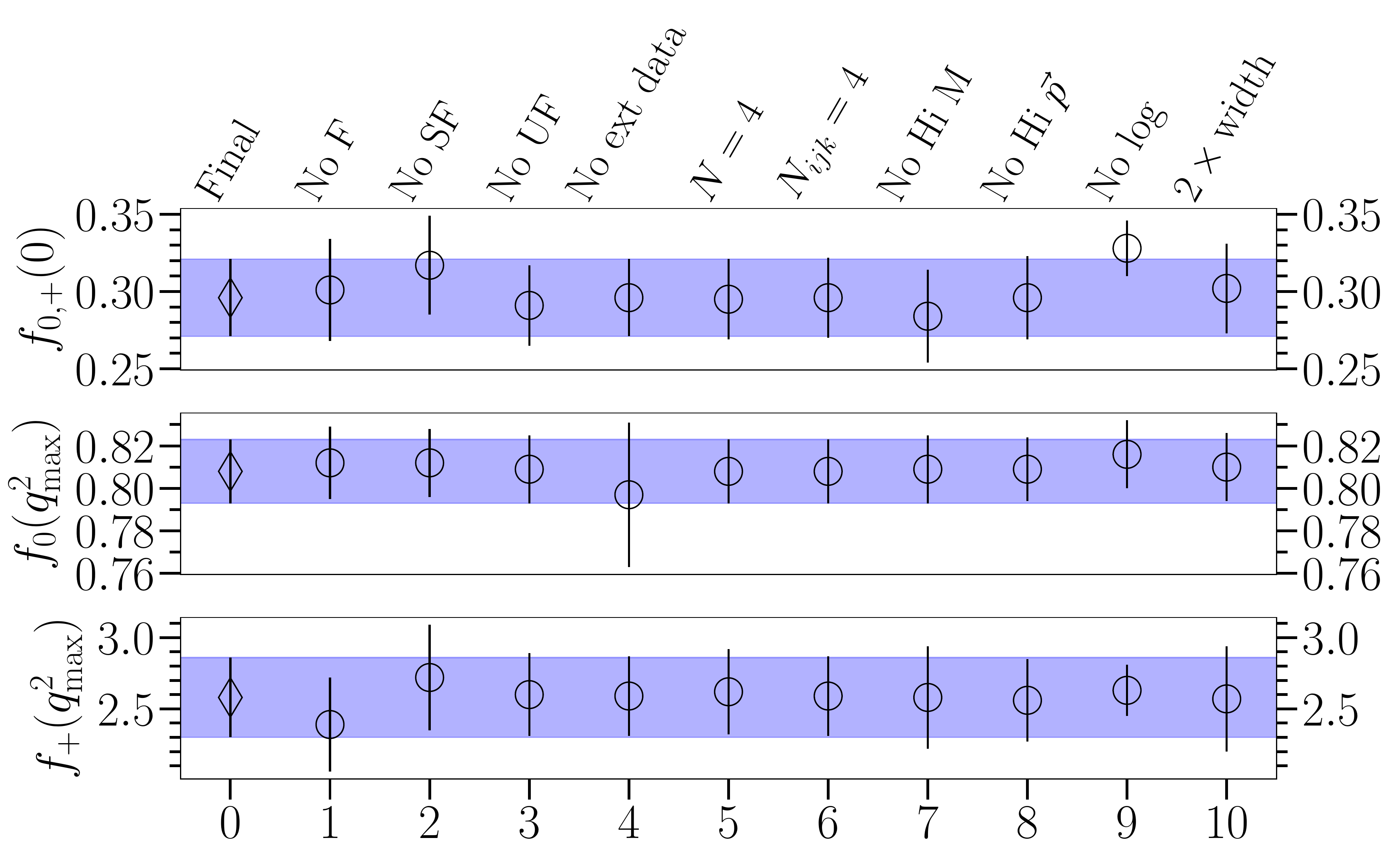}
  \caption{
  Stability tests of the fit of the form factors $f_{0,+}(0)$, $f_0(q^2_{\rm max})$ and $f_+(q^2_{\rm max})$. 
  Test 0 is the final result, shown throughout by the blue band. 
  Tests 1, 2 and 3 are the results if the fine, superfine and ultrafine data are removed respectively. 
  Test 4 is the fit without the data point from~\cite{Colquhoun:2015mfa}. 
  Test 5 adds a cubic term in the $z$ expansion [Eq.~(\ref{Eq:zexpansion})]. 
  Test 6 shows the effect of extending the $i,j,k$ sum in Eq.~(\ref{Eq:an}). 
  Tests 7 and 8 remove the highest masses and momenta for all lattice spacings respectively. 
  Test 9 is without the log term in Eq.~(\ref{Eq:an}); here we find that $d_{i000}$ terms change to mimic the Taylor expansion of the log, and we require much larger priors ($0\pm 5$) to account for this. 
  Test 10 shows the effect of doubling the width of all $d_{ijkn}$ priors. 
  We see that our extrapolation is stable to all of the above modifications. 
  Increasing the prior widths decreases the GBF, giving us confidence our priors are chosen conservatively.
  }
  \label{Fig:ExtrapStab}
\end{figure}
In Figs.~\ref{Fig:f0nopoleinz} and~\ref{Fig:fplusnopoleinz} we show our lattice data in $z$ space, as well as the results of the fit at the physical point for each form factor. In both cases these are plotted with their respective poles removed. 
We see very little $z$ dependence in the $f_0$ case, which we can also infer from our $a_1^0$ and $a_2^0$ values (Table~\ref{tab:ancoefficients}), both of which are consistent with zero. 
In contrast, $f_+$ displays a negative linear $z$ dependence, again clear in the expansion coefficients. 
Both of these trends are similar to the findings in~\cite{McLean:2019qcx}. 
Both cases have large errors in some ultrafine data, which simply arises from lack of statistics on the very computationally expensive ultrafine configurations. 

The lowest masses on each set correspond approximately to the physical charm mass, and we can see in Fig.~\ref{Fig:f0nopoleinz} that these points lie on top of each other, indicating that lattice artefacts such as discretisation errors are small at this mass. 
Other masses that are approximately equal are the ultrafine $am_h = 0.45$ and superfine $am_h = 0.6$, the ultrafine $am_h = 0.6$ and superfine $am_h = 0.8$, and the superfine $am_h = 0.45$ and fine $am_h = 0.683$. 
By comparing these values in Fig.~\ref{Fig:f0nopoleinz} we can see that, whilst lattice artefacts become slightly more significant above the charm mass, they are still small, and that the heavy mass dependence itself is what dominates this plot. 
The picture is less clear in Fig.~\ref{Fig:fplusnopoleinz} because of larger statistical errors, but it appears to be similarly dominated by heavy quark mass dependence.

We verify our results for the form factors at the physical point are stable with respect to reasonable variations of the fit by modifying the fit as illustrated in Fig.~\ref{Fig:ExtrapStab} and discussed in the caption.
The fit is stable under these variations, suggesting associated systematic uncertainties are adequately accounted for.

\subsubsection{Form factor error budget}
\begin{figure}
  \hspace{-12pt}
  \includegraphics[width=0.50\textwidth]{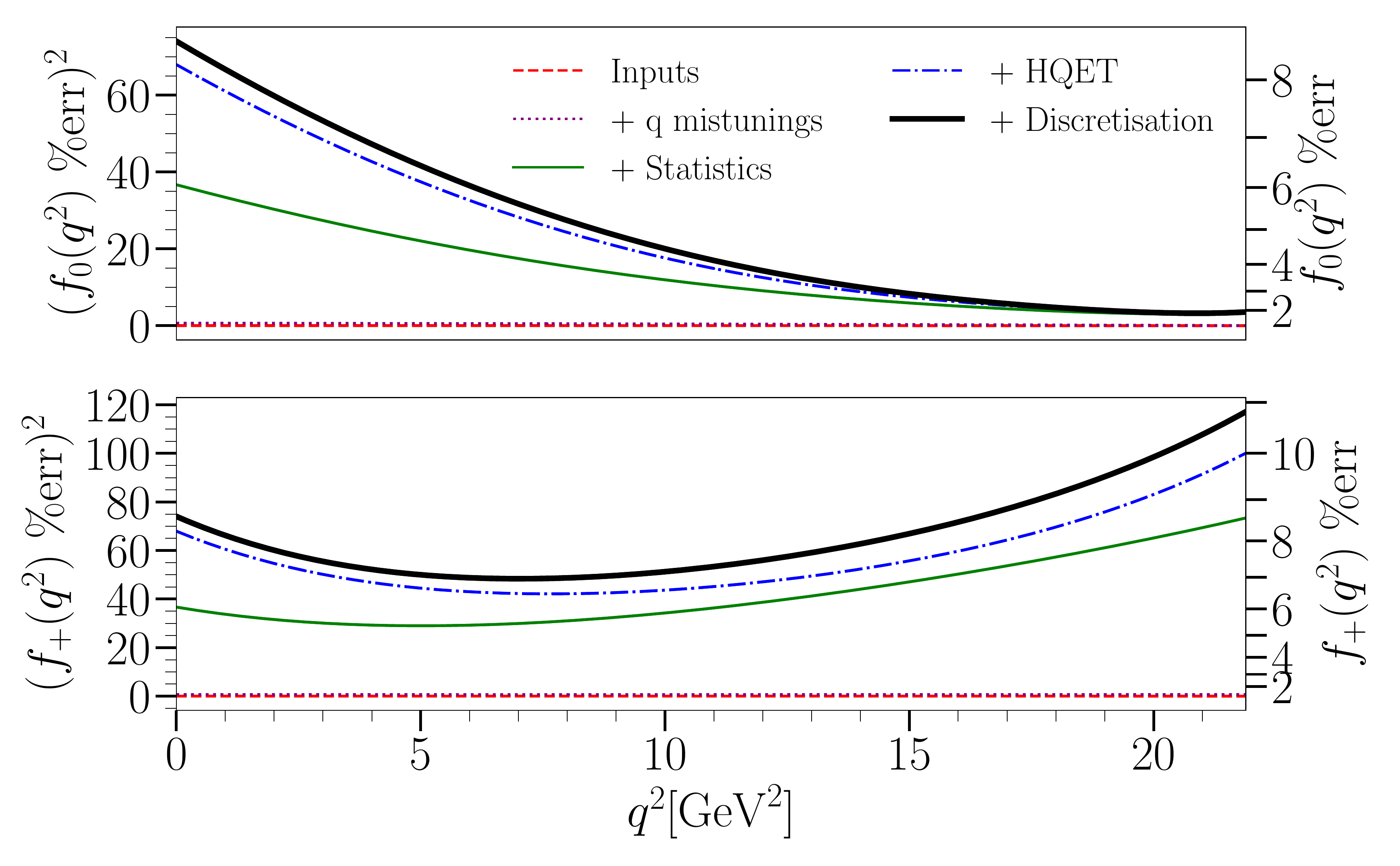}
  \caption{
  The contributions to the total percentage error (black line) of $f_0(q^2)$ (top) and $f_+(q^2)$ (bottom) from different sources, shown as an accumulating error. 
  The red dashed line (``inputs") includes values for masses taken from the PDG ~\cite{PDG} and used in the fit as described above. 
  The purple dotted line (``$q$ mistunings") adds, negligibly, to the inputs the error contribution from the quark mistunings associated with $c$ fit parameters, whilst the solid green line (``statistics") further adds the error from our correlator fits. 
  The blue dot-dashed line (``HQET") includes the contribution from the expansion in the heavy quark mass, and, finally, the thick black line (``Discretisation"), the total error on the form factor, also includes the discretisation errors. 
  The percentage variance adds linearly and the scale for this is given on the left-hand axis. 
  The percentage standard deviation, the square root of this, can be read from the scale on the right-hand side.
  }
  \label{Fig:f0fpluserr}
\end{figure}
Fig.~\ref{Fig:f0fpluserr} shows how the relative percentage error of each of the form factors builds up as contributions are added. 
This is plotted over the full $q^2$ range. We note that the error in the $f_0$ form factor shrinks with $q^2$, whilst that in $f_+$ grows. 
This is true even without the continuum data point from~\cite{Colquhoun:2015mfa}, so that statistical errors completely dominate $f_0(q^2_{\rm max})$. 
The vector form factor has a minimum error somewhere in between $0$ and $q^2_{\rm max}$, where our data are most densely distributed. 
This trend is similar in the scalar form factor if we remove the continuum data point which dominates the error at high $q^2$. 
We also note that the quark mistuning and input errors for both cases are small and almost independent of $q^2$, as we would expect. 
It is clear that our error is statistics dominated, which is a strong affirmation of the heavy HISQ method and nonperturbative current renormalisation, as well as of the suitability of our $z$ expansion. 
This also leaves the door open to a significant reduction in error, simply by increasing our statistics, particularly on the finest ensemble; a costly but straightforward exercise. 
We can see that, with sufficient computing time, errors could be reduced to 2\%--3\% across the full $q^2$ range for both the scalar and vector form factors. 

\section{Form Factor Results and Comparisons}\label{sec:results}
\begin{figure}
  \hspace{-12pt}
  \includegraphics[width=0.50\textwidth]{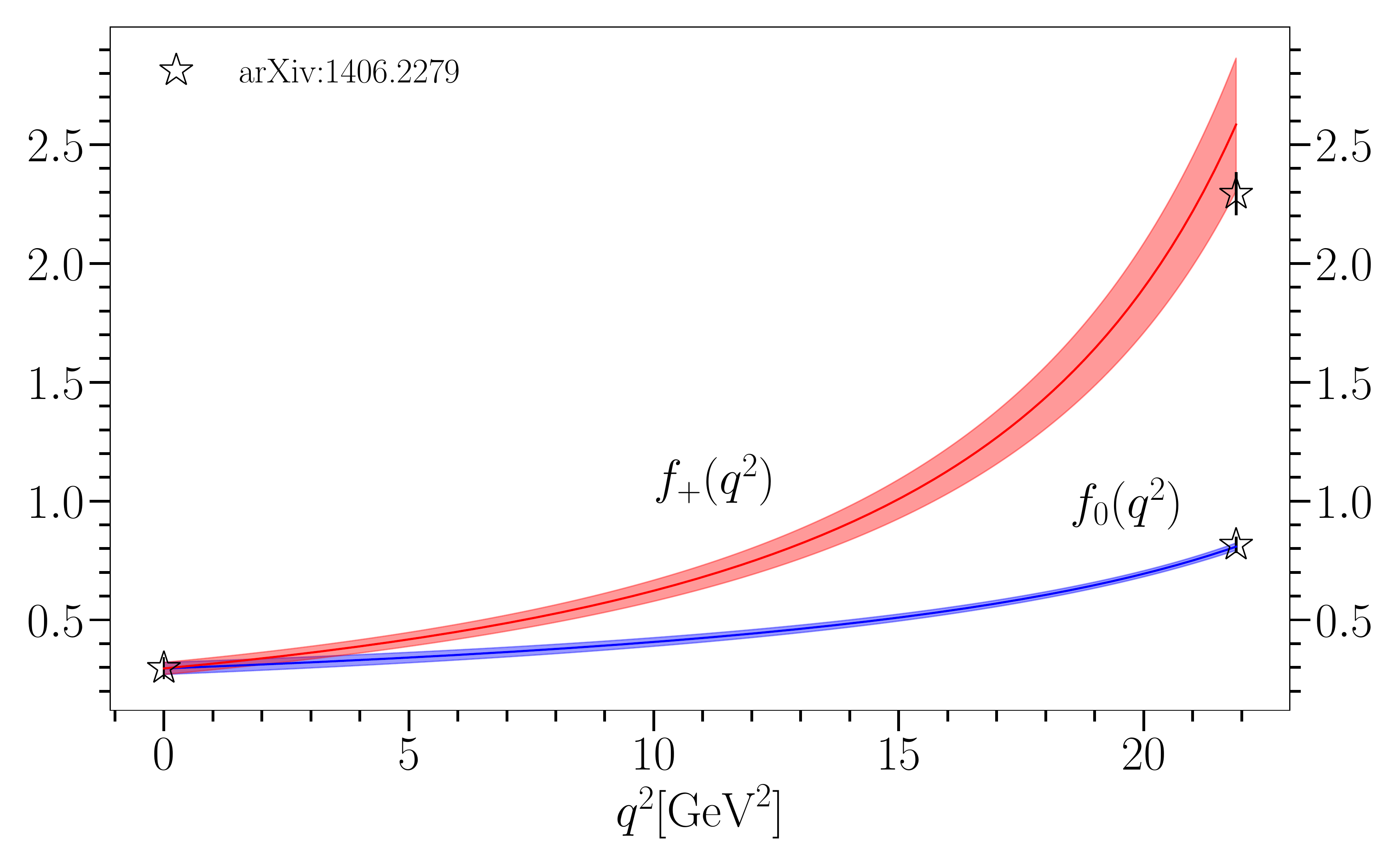}
  \caption{
  Final form factor results for $f_0(q^2)$ and $f_+(q^2)$. 
  Results from~\cite{Bouchard:2014ypa} at $q^2 = 0$ and $q^2 = q^2_{\rm max}$ are also shown.
  }
  \label{Fig:f0fplusinq}
\end{figure}
Our form factors at zero lattice spacing and physical quark mass are shown over the full physical $q^2$ range in Fig.~\ref{Fig:f0fplusinq}. 
We can compare these with $B_s \to \eta_s$ results from a lattice calculation that used NRQCD $b$ quarks given in the Appendix of~\cite{Bouchard:2014ypa}. 
We find the results to be in good agreement with an improvement in uncertainty across the $q^2$ range in the case of the $f_0$ form factor, and an improvement by a factor of 2 at $q^2=0$. 
The systematic uncertainties in the NRQCD calculation are dominated by the extrapolation to $q^2=0$ from high $q^2$ values close to zero recoil and the associated discretisation errors. 
The use of relatively coarse lattices in the NRQCD approach means that results are restricted to small daughter meson momentum. 
There is also a sizable systematic uncertainty from current renormalisation present in the NRQCD results. 
We do not have these sources of error here. 
Our result for $f_+(q^2_{\rm max})$ agrees to $1\sigma$ with the NRQCD value, but with significantly larger uncertainty. 
This is a region of $q^2$ space where our data have large statistical errors because of the way that $f_+$ is constructed from a temporal vector current in that limit. 
The differential rate for the decay vanishes rapidly toward $q^2_{\rm max}$ so it is the smaller values of $q^2$ at which we want to improve lattice QCD determination of the form factors and we have succeeded in doing this. 

\begin{figure*}
  \hspace{-30pt}
  \hspace{0.21in}
  \includegraphics[width=1.02\textwidth]{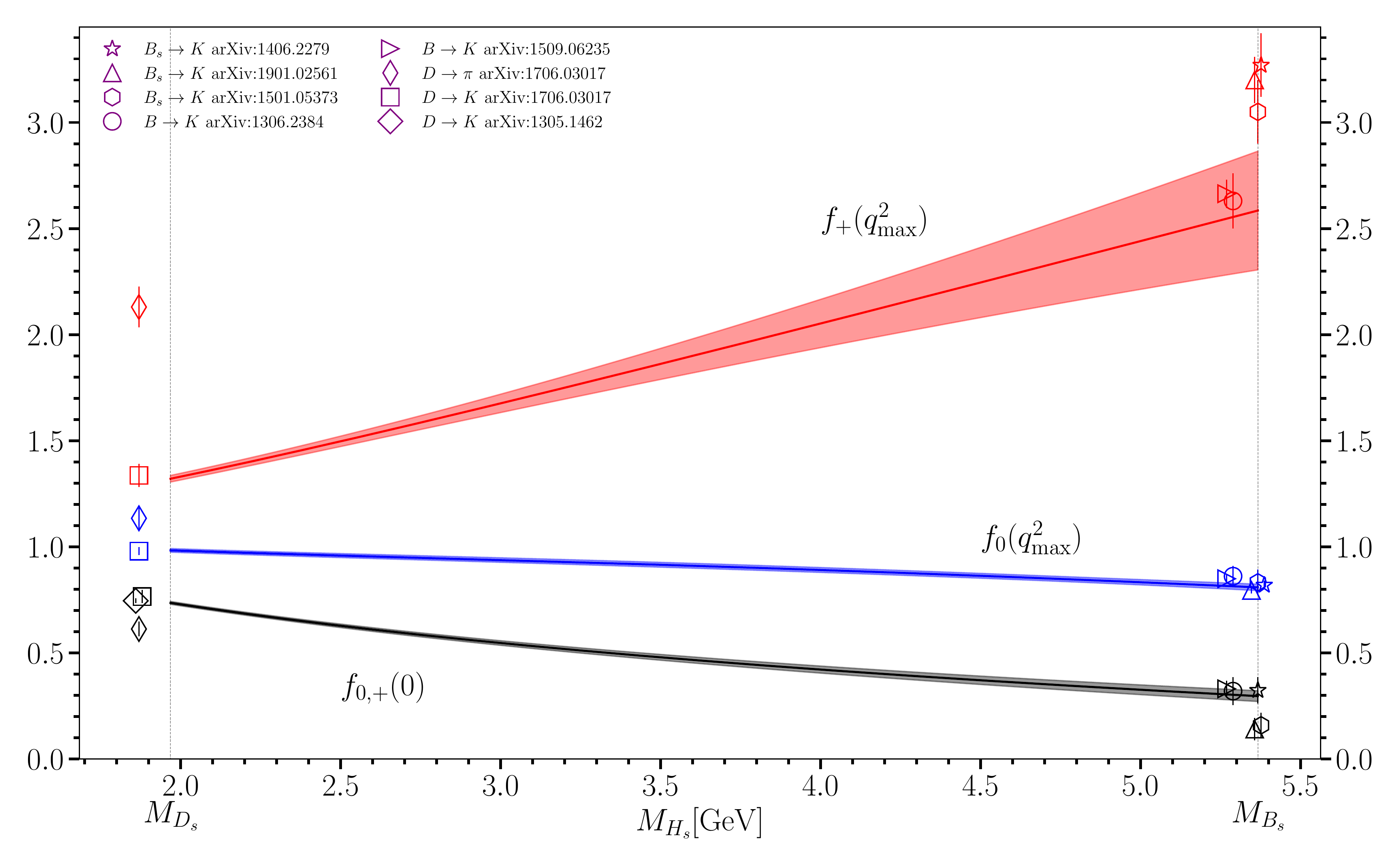}
  \caption{
  The form factors $f_{0,+}(0)$, $f_0(q^2_{\rm max})$ and $f_+(q^2_{\rm max})$ over the range of heavy masses from the physical $D_s$ to the physical $B_s$. 
  Results are included for $f_{0,+}(0)$, $f_{0}(q^2_{\rm max})$ and $f_{+}(q^2_{\rm max})$ (in their respective colours) for several other decays related by SU(3) flavour symmetry~\cite{Koponen:2013tua,Lattice:2015tia,Bouchard:2013eph,Bouchard:2014ypa,Lubicz:2017syv}. 
  Data points are plotted at the $x$ axis values corresponding to their physical heavy meson mass, not the mass that would result from their heavy quark and a strange quark (which would put them all at $M_{D_s}$ or $M_{B_s}$). 
  In the case of $M_B$ and $M_D$ some of the points are offset slightly either side of the mass for clarity.
  }
  \label{Fig:f0f0fplusinmh}
\end{figure*}
%

\subsection{Comparisons testing SU(3) flavour and heavy quark symmetries}
While the $B_s \to \eta_s$ decay does not correspond to a physical process, it is related to a host of physical decays via combinations of SU(3) flavour and heavy quark symmetry.
In this section, we evaluate these symmetries by comparing to published results for symmetry-related decays.

Fig.~\ref{Fig:f0f0fplusinmh} shows the effect of changing heavy quark mass over the full range of $M_{H_s}$ from the physical $M_{D_s}$ to the physical $M_{B_s}$, for both form factors at $q^2=0$ [recall that $f_+(0) = f_0(0)$] and at maximum physical $q^2$. 
Our use of a range of heavy masses from the physical charm to the physical bottom allows for good control of this heavy mass dependence. 
The uncertainty at the lighter end is particularly small, as all three ensembles had a physical charm mass data point, whereas only set 3 was fine enough to give data at the physical bottom mass. 
$f_{0,+}(0)$, $f_0(q^2_{\rm max})$ and $f_+(q^2_{\rm max})$ are converging as $M_{H_s}$ is reduced and one can imagine them meeting if extrapolated in mass below $M_{D_s}$ to $M_{\eta_s}$
That point would correspond to the $\eta_s\to\eta_s$ decay, where only $q^2=0$ is kinematically allowed and we expect $f_+ = f_0 = 1$.
A similar effect was seen in~\cite{McLean:2019qcx}.

Previous lattice QCD results for other decay processes related by SU(3) flavour symmetry are included in Fig.~\ref{Fig:f0f0fplusinmh} in the same colour labeling system. 
We see very good agreement with the $D \to K$ and $B \to K$ decays for both form factors at both ends of the $q^2$ range, suggesting that the mass of the spectator quark has almost no effect on the form factors, and supporting our use of $B_s \to \eta_s$ to test the viability of a $B \to K$ calculation. 
$B_s \to K$ data show good agreement for $f_0$ but $f_+(q^2_{\rm max})$ is in slight tension.
This suggests, as expected, that the form factors are much more sensitive to SU(3) flavour symmetry breaking in the daughter quark in the transition than in the spectator quark. 
This is further supported by the $D \to \pi$ results, which are in poor agreement with our $D_s \to \eta_s$ form factors across the board. 
$B \to \pi$ results are in even worse agreement and are not included in the plot.
This implies that symmetry breaking in the light daughter quark becomes even more important as the heavy parent quark becomes heavier.

\subsection{Tests of HQET}
%
\begin{figure}
  \hspace{-15pt}
  \includegraphics[width=0.50\textwidth]{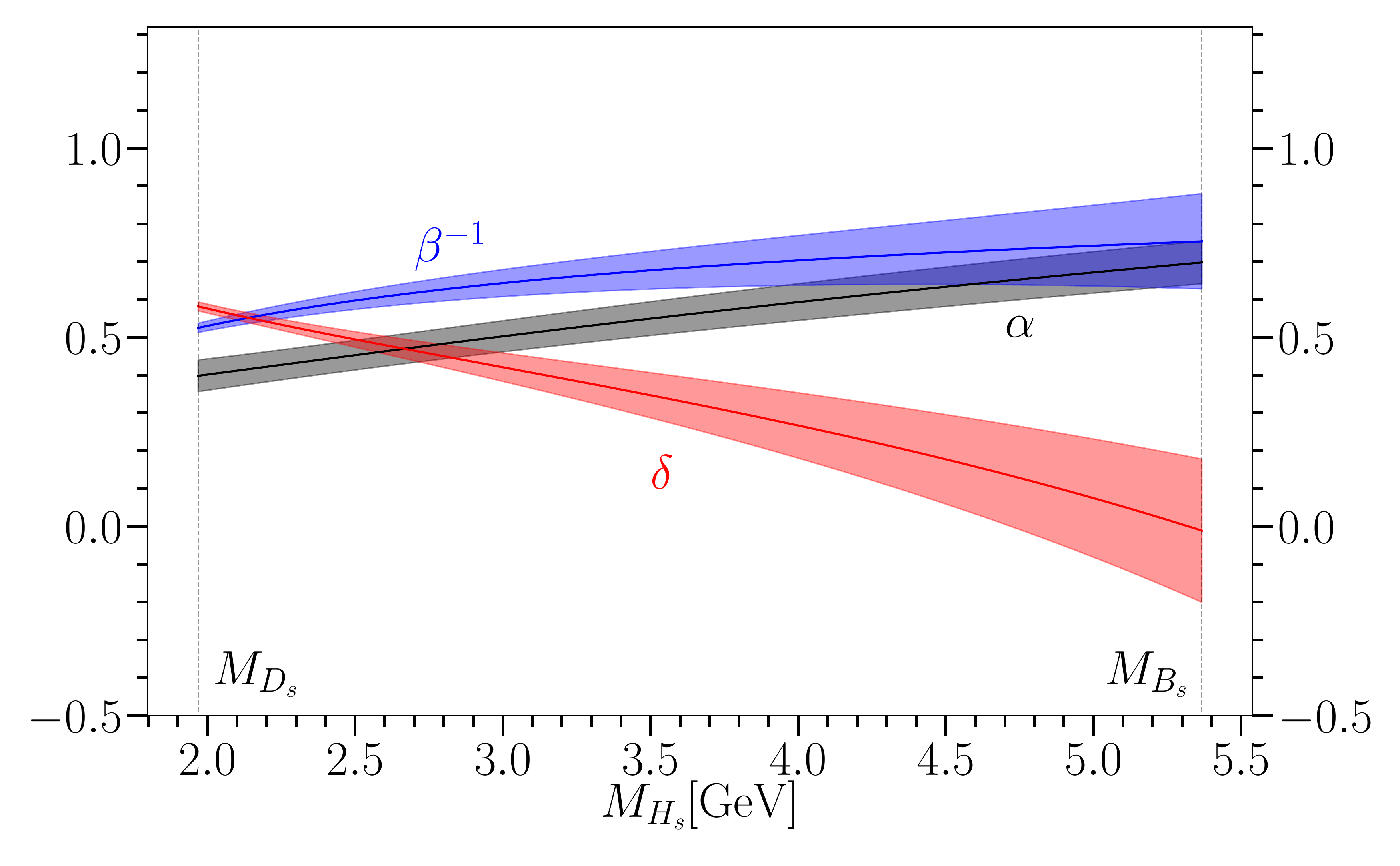}
  \caption{
  The quantities $\alpha$, $\beta^{-1}$ and $\delta$, defined in Eqs.~(\ref{alpha}),~(\ref{delta}) and~(\ref{beta}), over the range of heavy masses from the physical $D_s$ to the physical $B_s$.
  }
  \label{Fig:betadelta}
\end{figure}

That we are able to evaluate our form factors over the full range $m_c \leq m_h \leq m_b$ means we are in a unique position to test predictions of HQET.
One such set of predictions relates to the characterisation of form factor shape.
The quantities $\alpha$, $\delta$ and $\beta^{-1}$ are used to describe the shape of the form factors in HQET~\cite{Hill:2005ju,Hill:2006ub}. 
The latter two of these are related to the slope of the form factors at $q^2=0$ and the first to the value at high $q^2$:
\begin{align}
  \frac{1}{1-\alpha} &= \frac{1}{M^2_{H^*_s}}\mathrm{Res}_{q^2=M^2_{H^*_s}}\frac{f_+(q^2)}{f_+(0)}, \label{alpha} \\
  \delta &= 1 - \frac{M_{H_s}^2-M_{\eta_s}^2}{f_+(0)}\Bigg(\frac{df_+}{dq^2}\biggr\rvert_{q^2=0} - \frac{df_0}{dq^2}\biggr\rvert_{q^2=0}\Bigg), \label{delta} \\
  \frac{1}{\beta} &= \frac{M_{H_s}^2-M_{\eta_s}^2}{f_+(0)}\frac{df_0}{dq^2}\biggr\rvert_{q^2=0}. \label{beta}
\end{align}
Fig.~\ref{Fig:betadelta} shows our results for these quantities, plotted across the full range of heavy masses from $c$ to $b$ using as the $x$ axis the mass of the heavy-strange pseudoscalar meson. 
Our results for $\alpha$ and $\beta$ are qualitatively in agreement with expectations from HQET~\cite{Hill:2005ju} with $\alpha$ and $\beta$ close to one at the heaviest masses and differing further from one as the heavy quark mass falls. 
Our results are accurate enough that they could be used to constrain scaling laws in the mass from other theoretical approaches. 
We see that $\delta$ is close to zero at the $B_s$ end of the plot but clearly nonzero at the $D_s$ end. 
We find values of $\alpha_{M_{B_s}} = 0.698(56)$, $\beta_{M_{B_s}} = 1.33(22)$, $\delta_{M_{B_s}} = -0.01(19)$, $\alpha_{M_{D_s}} = 0.398(42)$, $\beta_{M_{D_s}} = 1.905(45)$ and $\delta_{M_{D_s}} = 0.582(12)$.

\begin{figure}
  \hspace{-12pt}
  \includegraphics[width=0.50\textwidth]{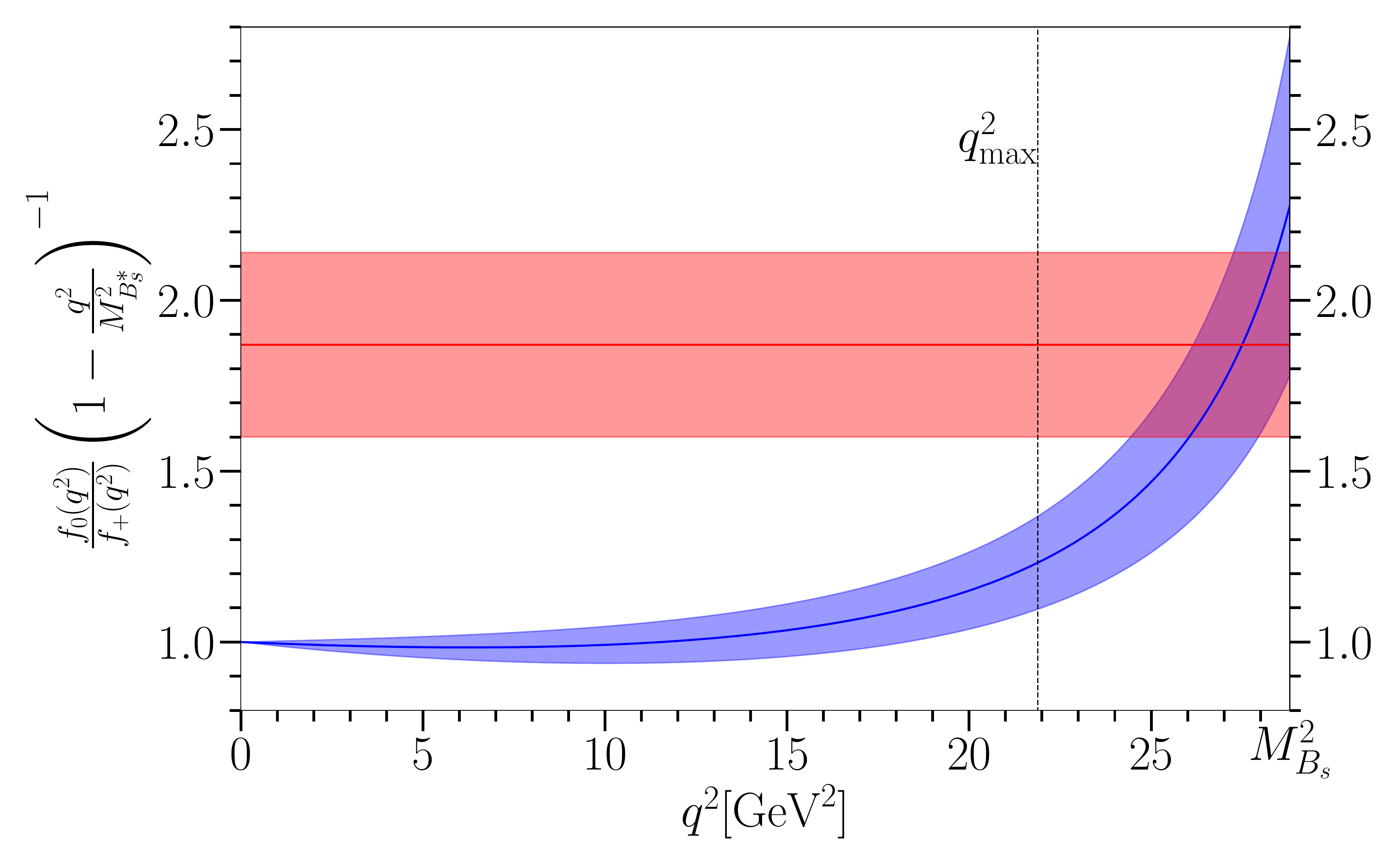}
  \caption{
  The form factor ratio, $\frac{f_0(q^2)}{f_+(q^2)}\Big(1-\frac{q^2}{M^2_{B_s^*}}\Big)^{-1}$ over the range $0 \leq q^2\leq M^2_{B_s}$ (blue band), as compared with the HQET expectation in the limit $q^2\to M^2_{B_s}$ (red band), defined in Eq.~(\ref{Eq:HQETrat}).
  }
  \label{Fig:HQETrat}
\end{figure}
The form factor ratio $\frac{f_0(q^2)}{f_+(q^2)}\Big(1-\frac{q^2}{M^2_{B_s^*}}\Big)^{-1}$ is shown in Fig.~\ref{Fig:HQETrat}, where it is compared with the HQET expectation~\cite{Burdman:1993es}
\begin{equation}\label{Eq:HQETrat}
  \lim_{q^2\to{}M^2_{B_s}}\frac{f_0(q^2)}{f_+(q^2)}\Big(1-\frac{q^2}{M^2_{B_s^*}}\Big)^{-1} = \Big(\frac{f_{B_s}}{f_{B^*_s}}\Big)\frac{1}{g_{B^*_sB_s\eta_{s}}}.
\end{equation}
This is included in~\cite{Burdman:1993es} as a $B \to \pi$ expectation; to test it here in $B_s \rightarrow \eta_s$ we replace $B$ with $B_s$. 
We take the ratio of decay constants $\frac{f_{B^*_s}}{f_{B_s}} = 0.953(23)$~\cite{Colquhoun:2015oha}. 
No difference is visible in this ratio between $B_s$ and $B$ in~\cite{Colquhoun:2015oha}. 
We take the coupling $g_{B^*_s B_s \eta_s} \approx g_{B^*B\pi} = 0.56(8)$~\cite{Flynn:2015xna}, because again the light quark mass dependence seen in~\cite{Flynn:2015xna} is mild. 
This leads us to expect little impact from SU(3) flavour symmetry breaking in our test of Eq.~(\ref{Eq:HQETrat}). 
This is also consistent with our observation in Fig.~\ref{Fig:f0f0fplusinmh} that SU(3) flavour symmetry breaking effects in the daughter quark affect both $f_0$ and $f_+$ at large $q^2$, and so there will be some cancellation of the effects in their ratio. 
Fig.~\ref{Fig:HQETrat} shows reasonable agreement with Eq.~(\ref{Eq:HQETrat}) in the limit $q^2\to{}M^2_{B_s}$, as is found for $B\to\pi$ in~\cite{Lattice:2015tia}. 

\begin{figure}
\hspace{-12pt}
\includegraphics[width=0.50\textwidth]{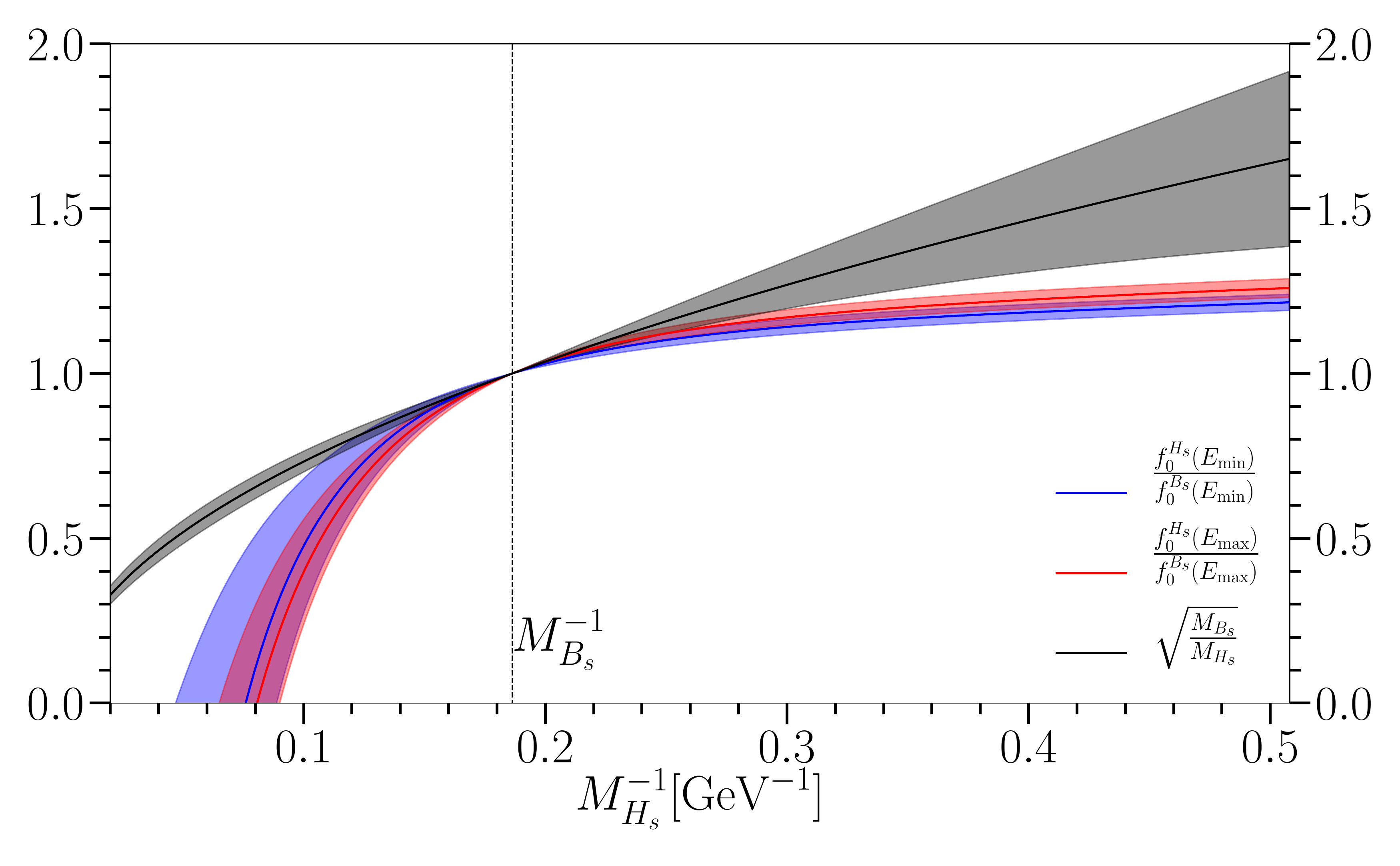}
\caption{
The form factor ratio $\frac{f_0^{H_s}(q^2(E))}{f_0^{B_s}(q^2(E))}$ evaluated at $\eta_s$ energy $E=E_{\rm min} = M_{\eta_s} = 0.6885(22)$\,GeV (blue line and error band) and at $E_{\rm max}$ corresponding to the largest energy available to the $\eta_s$ in a $D_s$ decay (red line and error band). 
Both ratios are plotted over a range of inverse heavy meson masses up to $M^{-1}_{D_s}$. 
The black dashed line marks $M^{-1}_{H_s} = M^{-1}_{B_s}$. 
Results are compared with the expectation of $\sqrt{\frac{M_{B_s}}{M_{H_s}}}$~\cite{Hill:2005ju}, given by the black band (see text).
}
\label{Fig:BothHillrat}
\end{figure}

Fig.~\ref{Fig:BothHillrat} tests the relationships between form factors for a changing initial state but fixed final state with a fixed energy. 
In~\cite{Hill:2005ju} it is shown that the $f_0$ form factor for a pseudoscalar heavy meson decay to a pseudoscalar light meson at fixed energy is inversely proportional to the square root of the heavy meson mass. 
This scaling should work both at small energy, close to zero recoil, and also at large energy, high recoil. 
In~\cite{Hill:2005ju} this is used to compare $B \rightarrow \pi$ and $D \rightarrow \pi$ decay. 
Here we compare $B_s \rightarrow \eta_s$ to $H_s \rightarrow \eta_s$ for variable $H_s$ mass from $D_s$ upward. 

Fig.~\ref{Fig:BothHillrat} compares $f_0(H_s \rightarrow \eta_s(E)) / f_0(B_s \rightarrow \eta_s(E))$ to the expectation $\sqrt{M_{B_s}/M_{H_s}}$ given by the black line.
We include an error in the HQET expectation from higher-order HQET terms of $\pm \sqrt{\frac{M_{B_s}}{M_{H_s}}}\Lambda_{\rm QCD} |M^{-1}_{H_s} - M^{-1}_{B_s}|$. 
Results are shown at two energies: the blue line and error band give results at zero recoil ($E_{\rm min} = M_{\eta_s}$) and the red line and error band give results at a higher energy, the maximum energy available to an $\eta_s$ in a $D_s$ decay [$E_{\rm max} = (M^2_{D_s} + M^2_{\eta_s}) / 2M_{D_s} = 1.105$\,GeV].
Our results at both energies are flatter than the $\sqrt{1/M_{H_s}}$ expectation, indicating that sizable corrections are needed to this expectation to describe the physical behaviour. 
This is reminiscent of results for the decay constant of heavy-strange pseudoscalar mesons in that it does not vary so strongly with mass as predicted;~\cite{McNeile:2011ng} shows that this decay constant only changes by 9.4(1.4)\% over the range from $c$ to $b$ when the leading-order HQET behaviour is as $\sqrt{1/M_{H_s}}$, i.e. a 65\,\% change.

\begin{figure}
\hspace{-12pt}
\includegraphics[width=0.50\textwidth]{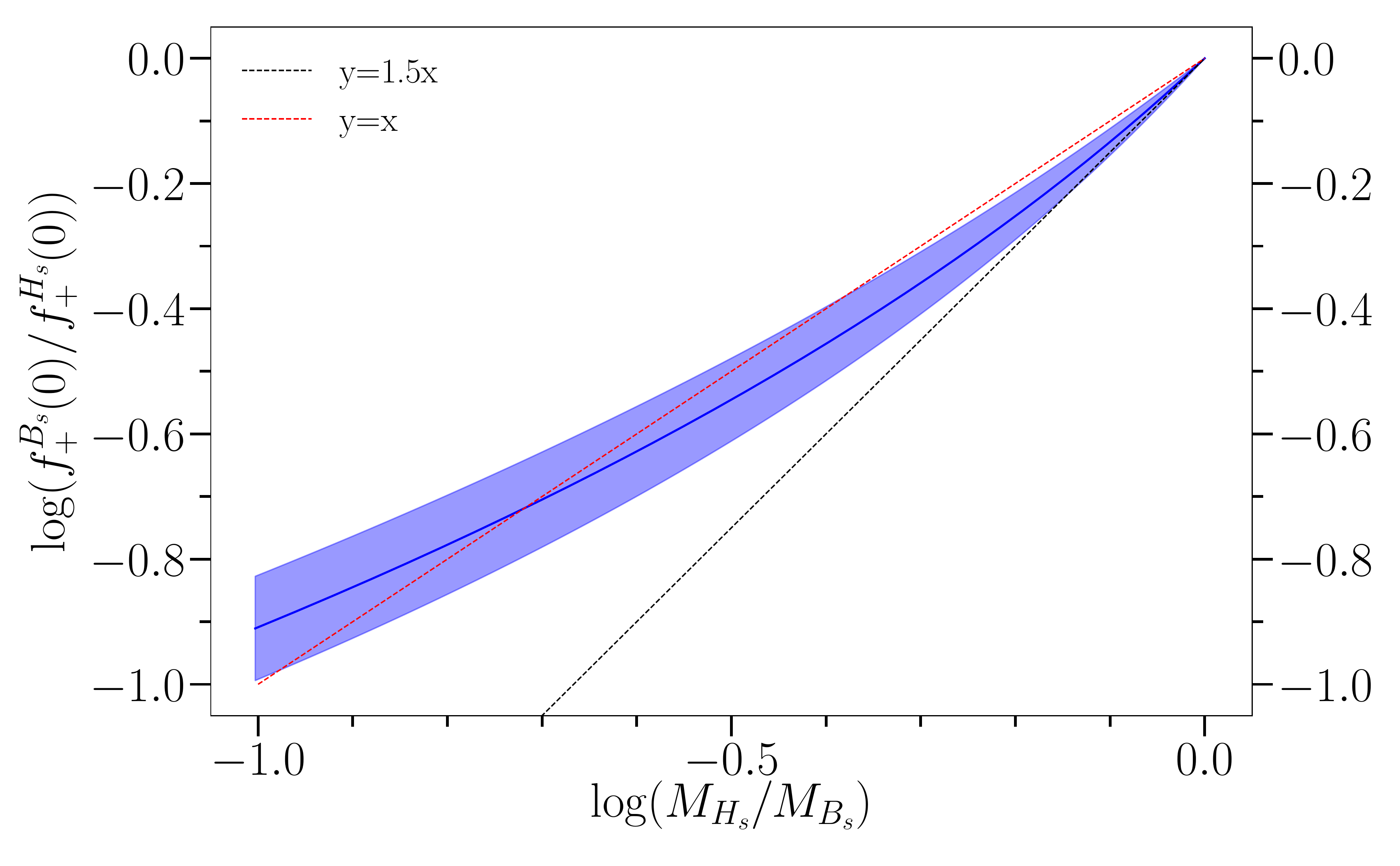}
\caption{
The form factor ratio $\frac{f_+^{B_s}(0)}{f_+^{H_s}(0)}$ plotted against the meson mass ratio $M_{H_s}/M_{B_s}$ in a log-log plot. 
Our results are shown as a blue curve with error band. 
The HQET expectation that the form factor ratio should depend on the 3/2 power of the mass ratio is shown as a black dashed line. 
In contrast, the red dashed line shows linear dependence on the mass. 
Results for the $D_s$ meson correspond to the left-hand end of the plot, $\log(M_{D_s}/M_{B_s}) = -1.003$. 
}
\label{Fig:Hillratlog}
\end{figure}

Finally, large-recoil scaling laws~\cite{Chernyak:1990ag,Becirevic:1999kt} give the prediction $\frac{f_+^{B_s}(0)}{f_+^{H_s}(0)} = (\frac{M_{H_s}}{M_{B_s}})^{3/2}$ at leading order. 
We examine this in Fig.~\ref{Fig:Hillratlog}, showing our results as a blue band and the HQET expectation as a black dashed line. 
We see that indeed the HQET expectation is borne out in the large heavy mass region close to the $b$. 
There are large corrections away from this region, however.
We find $\frac{f_+^{B_s}(0)}{f_+^{D_s}(0)} = 0.402(33)$ which is almost twice the size of $(\frac{M_{D_s}}{M_{B_s}})^{3/2} = 0.222$~\cite{PDG}. 

\section{Conclusions}\label{sec:conclusions}
We have performed the first calculation of form factors for a $b \rightarrow$ light quark transition in which we use our heavy-HISQ technique. 
This requires results at multiple values of the heavy quark mass on multiple sets of gluon field configurations with fine lattice spacing (going down to 0.045\,fm here) so that we can map out the heavy quark mass dependence of the form factors and obtain physical results for a heavy quark mass equal to that of the $b$. 
One advantage of this technique over previous calculations is that we can normalise the lattice currents completely nonperturbatively. 
Here we do this for the vector and scalar currents that give the vector and scalar form factors. 
This means that we can avoid sizable systematic errors from the one-loop matching of lattice currents to continuum currents that is done, for example, for NRQCD $b$ quarks. 
A second advantage of the heavy-HISQ technique is that it enables us to cover the full range in $q^2$ of the decay rather than just values of $q^2$ close to zero recoil (low momentum for the daughter meson). 
This is possible because the accessible range in $q^2$ grows as the accessible range in heavy quark mass grows on finer lattices. 

As a stepping-stone toward a variety of physical decay processes, we have chosen to study first the unphysical process $B_s \rightarrow \eta_s$ here because this does not involve valence $u$ or $d$ quarks and the $s$ quark mass can be accurately tuned to its physical value on all of our gluon field configurations. 
We present our final form factor results in Fig.~\ref{Fig:f0fplusinq}.
The form factor values at the end points of the $q^2$ range are:
\begin{eqnarray}
f_{0,+}(0) &= 0.296(25) \nonumber\\
f_{0}(q^2_{\rm max}) &= 0.808(15)  \\
f_{+}(q^2_{\rm max}) &= 2.58(28). \nonumber
\end{eqnarray}
Our uncertainty for the form factor at the kinematically important point (for the differential rate) $q^2 = 0$ is 8\%. 
This is an improvement by a factor of 2 over earlier results that used NRQCD $b$ quarks and coarser lattices. 
The uncertainties of the NRQCD result were dominated by the extrapolation of lattice results from relatively high $q^2$ values to $q^2 = 0$, along with the associated discretisation effects, statistical errors and a current matching uncertainty of 3\%. 
Our error budget as a function of $q^2$ is given in Fig.~\ref{Fig:f0fpluserr} and is dominated by statistical errors that can be improved at the cost of additional computing resource, to 2\%--3\% over the full $q^2$ range. 

Although our results correspond to an unphysical process, $B_s \rightarrow \eta_s$ is related to physical processes through SU(3) flavour symmetry for the light quark. 
Because we have results for the range of heavy quark masses from $c$ to $b$ we can study this SU(3) symmetry breaking through comparison to previous lattice QCD results for the physical processes for both $B$ and $D$ decay. 
This is shown in Fig.~\ref{Fig:f0f0fplusinmh}. 
We find that SU(3) flavour symmetry breaking in the daughter quark in the transition affects the form factors increasingly as the parent quark gets lighter. 
In contrast, symmetry breaking in the spectator quark has very little effect. 

HQET expectations for the mass scaling behaviour of form factors for $h \rightarrow l$ decay should hold for $B_s \rightarrow \eta_s$ up to effects from the $s$ quark mass, which should be small. 
We show comparison to such expectations in Fig.~\ref{Fig:HQETrat}--\ref{Fig:Hillratlog}. 
The latter two show substantial corrections to the leading-order HQET behaviour are present. 

Our results provide further evidence that the heavy-HISQ approach is an improved method for calculating hadronic form factors for semileptonic decays involving heavy quarks. 
This leads us to conclude that a heavy HISQ calculation of form factors for a physical $b\rightarrow s$ process, $B\to{}K\ell^+\ell^-$ will be able to improve upon the previous errors in~\cite{Bailey:2015dka,Bouchard:2013pna}. 
An accurate determination of the renormalisation of the lattice tensor current~\cite{Hatton:2020vzp}, possible with HISQ quarks, will allow us to improve the determination of the tensor form factor for that process as well. 
Our results are also encouraging for similar calculations involving $b \to l$ decays, such as $B \to \pi$ and $B_s \to K$, enabling improvement in the determination of the CKM element $V_{ub}$ when combined with experimental results. 

\subsection*{Acknowledgements}
We are grateful to the MILC Collaboration for the use of  their  configurations  and  their  code. 
We would also like to thank J. Harrison and G. P. Lepage for useful discussions and A. T. Lytle and J. Koponen for generating propagators. 
Computing was done on the Cambridge Service for Data Driven Discovery (CSD3) supercomputer, part of which is operated by the University of Cambridge Research Computing Service on behalf of the United Kingdom Science and Technology Facilities Council (STFC) Distributed Research utilising Advanced Computing (DiRAC) High Performance Computing (HPC) Facility.  
The DiRAC component of CSD3 was funded by BEIS via STFC capital grants and is operated by STFC operations grants. 
We are grateful to the CSD3 support staff for assistance. 
Funding for this work came from STFC. 

\begin{appendix}
\section{Correlator fit results}\label{sec:appendix}

\begin{table*}
  \caption{
  Results from fits to correlators on set 1. 
  For each heavy quark mass there are five values for the $\eta_s$ momentum, giving five different values for $q^2$. 
  For each of these values we give the ground state energy of the $\eta_s$, as well as the two current matrix elements (the matrix element for the vector is given before renormalisation with $Z_V$). 
  The final two columns give the values for $f_0(q^2)$ and $f_+(q^2)$, determined using Eqs.~(\ref{Eq:vec}) and~(\ref{Eq:sca}).
  }
  \begin{center} 
    \begin{tabular}{c c c c c c c c c}
      \hline
      Set &  $am^{\text{val}}_{h}$& $aM_{H_s}$ &$(aq)^2$&$aE_{\eta_s}$ &$Z_{\rm disc\,}a\!\bra{\eta_s}S\ket{H_s}$ & $Z_{\rm disc\,} a\!\bra{\eta_s}V^0\ket{\hat{H}_s}$  &$f_0(q^2)$&$f_+(q^2)$\\ [0.5ex]
      \hline
      \hline 
      1&0.449&0.90084(11)&0.34436(15)&0.314015(89)&1.6978(32)&1.1830(27)&0.9798(18)&\\ [1ex]
      &&&0.32936(15)&0.322342(87)&1.6715(33)&1.1635(28)&0.9646(19)&1.282(90)\\ [1ex]
      &&&0.22140(13)&0.382266(73)&1.5117(34)&1.0473(29)&0.8724(19)&1.0421(88)\\ [1ex]
      &&&0.043910(98)&0.480778(58)&1.3129(60)&0.9137(52)&0.7576(35)&0.7811(38)\\ [1ex]
      &&&$-0.059421(83)$&0.538131(52)&1.219(18)&0.857(14)&0.704(10)&0.6776(98)\\ [1ex]
      \hline 
      1&0.566&1.03355(13)&0.51773(20)&0.314015(89)&1.7762(36)&1.2902(35)&0.9679(19)&\\ [1ex]
      &&&0.50052(20)&0.322342(87)&1.7482(37)&1.2686(36)&0.9527(20)&1.38(15)\\ [1ex]
      &&&0.37665(18)&0.382266(73)&1.5797(38)&1.1402(36)&0.8608(20)&1.114(15)\\ [1ex]
      &&&0.17301(14)&0.480778(58)&1.3707(65)&0.9923(59)&0.7470(35)&0.8278(56)\\ [1ex]
      &&&0.05446(13)&0.538131(52)&1.273(19)&0.930(15)&0.694(10)&0.714(11)\\ [1ex]
      \hline 
      1&0.683&1.16007(14)&0.71582(27)&0.314015(89)&1.8494(40)&1.3880(46)&0.9570(20)&\\ [1ex]
      &&&0.69649(26)&0.322342(87)&1.8199(42)&1.3643(47)&0.9418(21)&1.47(22)\\ [1ex]
      &&&0.55746(23)&0.382266(73)&1.6436(42)&1.2248(45)&0.8505(21)&1.189(24)\\ [1ex]
      &&&0.32890(20)&0.480778(58)&1.4255(69)&1.0639(69)&0.7377(35)&0.8758(87)\\ [1ex]
      &&&0.19583(18)&0.538131(52)&1.324(20)&0.997(17)&0.685(10)&0.752(13)\\ [1ex]
      \hline 
      1&0.8&1.28117(16)&0.93539(33)&0.314015(89)&1.9194(44)&1.4795(60)&0.9485(21)&\\ [1ex]
      &&&0.91406(33)&0.322342(87)&1.8885(45)&1.4540(60)&0.9333(21)&1.58(31)\\ [1ex]
      &&&0.76051(30)&0.382266(73)&1.7055(44)&1.3041(56)&0.8428(21)&1.270(35)\\ [1ex]
      &&&0.50809(26)&0.480778(58)&1.4790(72)&1.1309(79)&0.7309(35)&0.927(13)\\ [1ex]
      &&&0.36113(24)&0.538131(52)&1.375(20)&1.059(19)&0.679(10)&0.792(17)\\ [1ex]
      \hline
      \hline
    \end{tabular}
  \end{center}
  \label{tab:finefitresults}
\end{table*}

\begin{table*}
  \caption{Results from fits to correlators on set 2. For each heavy quark mass there are five values for the $\eta_s$ momentum, giving five different values for $q^2$. For each of these values we give the ground state energy of the $\eta_s$, as well as the two current matrix elements (the matrix element for the vector is given before renormalisation with $Z_V$). The final two columns give the values for $f_0(q^2)$ and $f_+(q^2)$, determined using Eqs.~(\ref{Eq:vec}) and~(\ref{Eq:sca}).}
  \begin{center} 
    \begin{tabular}{c c c c c c c c c}
      \hline
      Set &  $am^{\text{val}}_{h}$& $aM_{H_s}$ &$(aq)^2$&$aE_{\eta_s}$ &$Z_{\rm disc\,}a\!\bra{\eta_s}S\ket{H_s}$ & $Z_{\rm disc\,}a\!\bra{\eta_s}V^0\ket{\hat{H}_s}$  &$f_0(q^2)$&$f_+(q^2)$\\ [0.5ex]
      \hline
      \hline 
      2&0.274&0.59142(13)&0.14776(11)&0.207020(84)&1.2075(65)&0.7870(47)&0.9859(51)&\\ [1ex]
      &&&0.095058(93)&0.251579(69)&1.0731(64)&0.6966(51)&0.8762(51)&1.041(23)\\ [1ex]
      &&&0.018658(73)&0.316169(55)&0.924(13)&0.600(11)&0.755(11)&0.778(12)\\ [1ex]
      &&&$-0.072133(52)$&0.392925(44)&0.815(30)&0.532(22)&0.666(24)&0.598(23)\\ [1ex]
      &&&$-0.151516(34)$&0.460038(38)&0.717(94)&0.492(73)&0.585(77)&0.491(68)\\ [1ex]
      \hline 
      2&0.450&0.80078(20)&0.35256(25)&0.207020(84)&1.3367(92)&0.9568(79)&0.9529(64)&\\ [1ex]
      &&&0.28119(23)&0.251579(69)&1.1845(88)&0.8433(81)&0.8444(61)&1.207(71)\\ [1ex]
      &&&0.17775(20)&0.316169(55)&1.016(16)&0.720(14)&0.724(11)&0.891(30)\\ [1ex]
      &&&0.05482(17)&0.392925(44)&0.890(34)&0.631(27)&0.634(24)&0.672(26)\\ [1ex]
      &&&$-0.05267(14)$&0.460038(38)&0.78(10)&0.576(85)&0.559(74)&0.535(69)\\ [1ex]
      \hline 
      2&0.6&0.96656(27)&0.57690(42)&0.207020(84)&1.432(12)&1.078(11)&0.9261(74)&\\ [1ex]
      &&&0.49077(39)&0.251579(69)&1.266(11)&0.948(10)&0.8190(71)&1.34(12)\\ [1ex]
      &&&0.36590(35)&0.316169(55)&1.083(18)&0.807(17)&0.700(12)&0.980(55)\\ [1ex]
      &&&0.21753(31)&0.392925(44)&0.945(37)&0.703(31)&0.611(24)&0.734(37)\\ [1ex]
      &&&0.08779(27)&0.460038(38)&0.84(11)&0.638(95)&0.541(72)&0.576(82)\\ [1ex]
      \hline 
      2&0.8&1.17473(36)&0.93646(71)&0.207020(84)&1.545(16)&1.217(14)&0.8972(88)&\\ [1ex]
      &&&0.83177(67)&0.251579(69)&1.365(15)&1.070(13)&0.7926(83)&1.50(20)\\ [1ex]
      &&&0.68002(62)&0.316169(55)&1.165(21)&0.910(19)&0.677(12)&1.095(92)\\ [1ex]
      &&&0.49968(57)&0.392925(44)&1.013(41)&0.787(35)&0.588(23)&0.819(59)\\ [1ex]
      &&&0.34200(52)&0.460038(38)&0.90(12)&0.71(11)&0.525(70)&0.64(12)\\ [1ex]

      \hline
      \hline
    \end{tabular}
  \end{center}
  \label{tab:superfinefitresults}
\end{table*}

\begin{table*}
  \caption{Results from fits to correlators on set 3. For each heavy quark mass there are five values for the $\eta_s$ momentum, giving five different values for $q^2$. For each of these values we give the ground state energy of the $\eta_s$, as well as the two current matrix elements (the matrix element for the vector is given before renormalisation with $Z_V$). The final two columns give the values for $f_0(q^2)$ and $f_+(q^2)$, determined using Eqs.~(\ref{Eq:vec}) and~(\ref{Eq:sca}).}
  \begin{center} 
    \begin{tabular}{c c c c c c c c c}
      \hline
      Set &  $am^{\text{val}}_{h}$& $aM_{H_s}$ &$(aq)^2$&$aE_{\eta_s}$ &$Z_{\rm disc\,}a\!\bra{\eta_s}S\ket{H_s}$ & $Z_{\rm disc\,}a\!\bra{\eta_s}V^0\ket{\hat{H}_s}$  &$f_0(q^2)$&$f_+(q^2)$\\ [0.5ex]
      \hline
      \hline 
      3&0.194&0.43980(32)&0.08164(20)&0.15407(17)&0.949(16)&0.591(14)&0.992(15)&\\ [1ex]
      &&&0.07172(19)&0.16535(15)&0.909(15)&0.564(15)&0.951(15)&1.25(27)\\ [1ex]
      &&&0.03985(16)&0.20159(13)&0.807(20)&0.499(17)&0.844(20)&0.963(49)\\ [1ex]
      &&&0.00198(13)&0.24464(10)&0.703(57)&0.438(43)&0.735(59)&0.739(60)\\ [1ex]
      &&&$-0.1598983(72)$&0.428673(59)&0.51(25)&0.34(18)&0.54(26)&0.39(21)\\ [1ex]
      \hline 
      3&0.45&0.74667(78)&0.35118(94)&0.15407(17)&1.136(38)&0.841(38)&0.922(30)&\\ [1ex]
      &&&0.33433(92)&0.16535(15)&1.087(37)&0.803(37)&0.883(28)&1.6(1.2)\\ [1ex]
      &&&0.28022(86)&0.20159(13)&0.962(36)&0.707(35)&0.781(28)&1.26(27)\\ [1ex]
      &&&0.21593(79)&0.24464(10)&0.834(73)&0.615(65)&0.677(59)&0.93(23)\\ [1ex]
      &&&$-0.05890(50)$&0.428673(59)&0.63(31)&0.50(27)&0.51(25)&0.49(24)\\ [1ex]
      \hline 
      3&0.6&0.9107(10)&0.5724(16)&0.15407(17)&1.225(57)&0.931(53)&0.887(40)&\\ [1ex]
      &&&0.5519(16)&0.16535(15)&1.172(54)&0.890(49)&0.849(38)&1.8(1.8)\\ [1ex]
      &&&0.4859(15)&0.20159(13)&1.037(49)&0.784(45)&0.751(35)&1.39(44)\\ [1ex]
      &&&0.4075(14)&0.24464(10)&0.898(83)&0.685(73)&0.651(60)&1.00(38)\\ [1ex]
      &&&0.0723(10)&0.428673(59)&0.68(34)&0.56(30)&0.49(25)&0.51(27)\\ [1ex]
      \hline 
      3&0.8&1.1177(13)&0.9285(25)&0.15407(17)&1.340(84)&1.056(74)&0.856(53)&\\ [1ex]
      &&&0.9033(25)&0.16535(15)&1.282(79)&1.009(69)&0.820(50)&2.1(2.8)\\ [1ex]
      &&&0.8223(24)&0.20159(13)&1.134(70)&0.889(61)&0.725(44)&1.58(72)\\ [1ex]
      &&&0.7261(23)&0.24464(10)&0.982(99)&0.775(87)&0.628(63)&1.13(65)\\ [1ex]
      &&&0.3147(18)&0.428673(59)&0.74(37)&0.63(34)&0.47(24)&0.54(35)\\ [1ex]

      \hline
      \hline
    \end{tabular}
  \end{center}
  \label{tab:ultrafinefitresults}
\end{table*}

\end{appendix}

\bibliography{BsEtaspaper}
\end{document}